\documentclass[twocolumn,showkeys,showpacs,preprintnumbers,prd,superscriptaddress,nofootinbib]{revtex4-2}
\usepackage{graphicx}
\usepackage{epsf}
\usepackage{bm}
\usepackage{amsmath}
\usepackage{amsfonts}
\usepackage{booktabs}
\usepackage{amssymb}
\usepackage{epstopdf}
\usepackage{natbib}
\usepackage{color}
\usepackage[dvipsnames,table]{xcolor}
\usepackage{physics}
\usepackage[colorlinks = true,
            linkcolor = blue,
            urlcolor  = blue,
            citecolor = blue,
            anchorcolor = blue]{hyperref}
\usepackage{verbatim}
\usepackage{lipsum}
\usepackage{float}
\usepackage{multirow}
\usepackage{soul}

\usepackage[capitalize]{cleveref}
\providecommand{\U}[1]{\protect\rule{.1in}{.1in}}

\newcommand{\be}{\begin{equation}}
\newcommand{\ee}{\end{equation}}
\newcommand{\bea}{\begin{eqnarray}}
\newcommand{\eea}{\end{eqnarray}}

\usepackage{tikz}

\newcommand{\pdlupbf}[1]{%
    \raisebox{-0.8ex}{%
        \begin{tikzpicture}[#1]%
            \node[] at (-2.75ex,0) {\textbf{PDL}};
            \draw [-stealth,line width=0.5](1.5ex,-0.85ex) -- (3.5ex,0.85ex);
            \draw[line width=0.65] (0,0) -- (5ex,0);
        \end{tikzpicture}%
    }%
}
\newcommand{\pdldbf}[1]{%
    \raisebox{-0.8ex}{
        \begin{tikzpicture}[#1]%
            \node[] at (-2.75ex,0) {\textbf{PDL}};
            \draw [-stealth,line width=0.5](1.5ex,0.85ex) -- (3.5ex,-0.85ex);
            \draw[line width=0.65] (0,0) -- (5ex,0);
        \end{tikzpicture}%
    }
}

\newcommand{\pdlup}[1]{%
    \raisebox{-0.8ex}{%
        \begin{tikzpicture}[#1]%
            \node[] at (-2.5ex,0) {PDL};
            \draw [-stealth](1.5ex,-0.85ex) -- (3.5ex,0.85ex);
            \draw (0,0) -- (5ex,0);
        \end{tikzpicture}%
    }%
}
\newcommand{\pdld}[1]{%
    \raisebox{-0.8ex}{
        \begin{tikzpicture}[#1]%
            \node[] at (-2.5ex,0) {PDL};
            \draw [-stealth](1.5ex,0.85ex) -- (3.5ex,-0.85ex);
            \draw (0,0) -- (5ex,0);%
        \end{tikzpicture}%
    }
}

\begin{document}

\title{Omnipotent dark energy: A phenomenological answer to the Hubble tension}

\author{Shahnawaz A. Adil}
\email{shazadil14@gmail.com}
\affiliation{Department of Physics, Jamia Millia Islamia, New Delhi-110025, India.}

\author{\"{O}zg\"{u}r Akarsu}
\email{akarsuo@itu.edu.tr}
\affiliation{Department of Physics, Istanbul Technical University, Maslak 34469 Istanbul, Turkey}

\author{Eleonora Di Valentino}
\email{e.divalentino@sheffield.ac.uk}
\affiliation{School of Mathematics and Statistics, University of Sheffield, Hounsfield Road, Sheffield S3 7RH, United Kingdom}

\author{Rafael C. Nunes}
\email{rafadcnunes@gmail.com}
\affiliation{Instituto de F\'{i}sica, Universidade Federal do Rio Grande do Sul, 91501-970 Porto Alegre RS, Brazil}
\affiliation{Divis\~ao de Astrof\'isica, Instituto Nacional de Pesquisas Espaciais, Avenida dos Astronautas 1758, S\~ao Jos\'e dos Campos, 12227-010, SP, Brazil}

\author{Emre \"{O}z\"{u}lker}
\email{ozulker17@itu.edu.tr}
\affiliation{Department of Physics, Istanbul Technical University, Maslak 34469 Istanbul, Turkey}

\author{Anjan A. Sen}
\email{aasen@jmi.ac.in}
\affiliation{Centre for Theoretical Physics, Jamia Millia Islamia, New Delhi-110025, India.}

\author{Enrico Specogna}
\email{especogna1@sheffield.ac.uk}
\affiliation{School of Mathematics and Statistics, University of Sheffield, Hounsfield Road, Sheffield S3 7RH, United Kingdom}

\begin{abstract}

This paper introduces the class of omnipotent dark energy (DE) models characterized by nonmonotonic energy densities that are capable of attaining negative values with corresponding equation of state parameters featuring phantom divide line (PDL) crossings and singularities. These nontrivial features are phenomenologically motivated by findings of previous studies that reconstruct cosmological functions from observations, and the success of extensions of $\Lambda$CDM, whose actual or effective DE density is omnipotent, in alleviating the observational discordance within $\Lambda$CDM. As an example, we focus on one embodiment of omnipotent DE, viz., the DE parametrization introduced in Di Valentino \textit{et al.} [Dark energy with phantom crossing and the H0 tension,
\href{
https://doi.org/10.3390/e23040404
}{Entropy 23, 404 (2021)}] (DMS20).
By updating and extending the datasets used in the original paper where it was introduced, we confirm the effectiveness of DMS20 in alleviating the observational discrepancies. Additionally, we uncover that its negative DE density feature, importance of which was not previously investigated, plays a crucial role in alleviating the tensions, along with the PDL crossing feature that the parametrization presupposes. In particular, we find that there is a positive correlation between the $H_0$ parameter and the scale~($a_{\rm p}$) at which DE density transitions from negative to positive, in agreement with previous studies that incorporate this transition feature. For our full dataset, the model yields $H_0=70.05\pm0.64$ (68\% CL) relaxing the $H_0$ tension with a preference of crossing to negative DE densities ($a_{\rm p}>0$ at 99\% CL), along with the constraint $a_m=0.922^{+0.041}_{-0.035}$ on the scale of the presupposed PDL crossing.

\end{abstract}

\maketitle

\section{Introduction}

The lambda cold dark matter ($\Lambda$CDM) scenario has emerged as the agreed cosmological model, widely accepted for its simplicity and remarkable ability to provide a coherent explanation for a wide array of astrophysical and cosmological observations. Nonetheless, despite its accomplishments, $\Lambda$CDM confronts various lingering challenges that hinder its capacity to comprehensively elucidate fundamental aspects regarding the structure and evolution of the Universe. These challenges revolve around three enigmatic components, of which we do not have a complete understanding: inflation, dark matter (DM), and dark energy (DE). 

With the advancement of precise observations~\cite{DiValentino:2020vhf}, it is expected that deviations from the baseline $\Lambda$CDM model will be revealed. In fact, discrepancies in the estimate of cosmological parameters have already surfaced in diverse observations, displaying varying degrees of statistical significance~\cite{Abdalla:2022yfr,DiValentino:2020vvd,DiValentino:2020srs,DiValentino:2020zio,Perivolaropoulos:2021jda}. The most prominent and statistically significant discrepancy is in the Hubble constant ($H_0$) between its value inferred from cosmic microwave background (CMB) data~\cite{Planck:2018vyg,ACT:2023kun,ACT:2020gnv,SPT-3G:2022hvq} assuming the $\Lambda$CDM model, and its direct measurements that rely on different methods and astrophysical observations~\cite{Huang:2019yhh,Blakeslee:2021rqi,Garnavich:2022hef,deJaeger:2022lit,Pesce:2020xfe,Kourkchi:2020iyz,FernandezArenas:2017dux,Birrer:2022chj,Moresco:2022phi,Ward:2022ghz,Tully:2022rbj,Sanders:2023jkl,Freedman:2021ahq,Anand:2021sum,Shajib:2023uig,Scolnic:2023mrv,Anderson:2023aga} (see also Refs.~\cite{Verde:2019ivm,Knox:2019rjx,DiValentino:2020zio,Jedamzik:2020zmd,DiValentino:2021izs,Perivolaropoulos:2021jda,Shah:2021onj,Abdalla:2022yfr,Kamionkowski:2022pkx} for reviews). Its statistical significance reaches more than $5\sigma$ for certain analyses~\cite{Riess:2021jrx,Riess:2022mme,Murakami:2023xuy} and this discrepancy, known as the $H_0$ tension, is widely considered to be a crisis. A less statistically significant but still intriguing discrepancy related to the amount of matter clustering is the $S_8$ tension~\cite{Abdalla:2022yfr,DiValentino:2020vvd}. Assuming the $\Lambda$CDM model, the CMB based constraints on the $S_8$ parameter are higher compared to those from late time measurements such as the weak-lensing estimates~\cite{DES:2021wwk,DES:2022ygi,KiDS:2020suj,Li:2023tui,Dalal:2023olq,Kilo-DegreeSurvey:2023gfr}. While these tensions and some further anomalies of $\Lambda$CDM~\cite{DiValentino:2020vhf,DiValentino:2020zio,DiValentino:2020vvd,DiValentino:2020srs,Perivolaropoulos:2021jda,Abdalla:2022yfr,DiValentino:2022oon,Annis:2022xgg} may arise from systematic errors, the fact that the above mentioned tensions have recurred across a multitude of investigations and the variety of the anomalies has increased over the years, when combined with the already ambiguous nature of DE and DM, points to flaws within the established cosmological framework and suggests the necessity for new physics to tackle these observational challenges and refine our understanding of the Universe.

A plethora of scenarios have been put forth in the literature with the aim of alleviating the cosmological tensions. However, to date, none of them have managed to convincingly and comprehensively resolve these issues (see for example Refs.~\cite{DiValentino:2020zio,DiValentino:2021izs,Perivolaropoulos:2021jda,Abdalla:2022yfr} and references therein). In particular, the failure of canonical/simple extensions of $\Lambda$CDM in satisfactorily addressing the observational tensions, indicate that the deviations from the standard $\Lambda$CDM model necessary for the resolution of its challenges may be highly nontrivial from the point of view of fundamental physics. For a deviation from the cosmological constant in the form of an actual or effective dynamical DE source, some examples of the studied nontrivial features are a density that attains negative values in the past accompanied by a singular equation of state (EoS) parameter~\cite{BOSS:2014hwf,Aubourg:2014yra,Sahni:2014ooa,DiValentino:2017rcr,Mortsell:2018mfj,Poulin:2018zxs,Capozziello:2018jya,Wang:2018fng,Akarsu:2019ygx,Dutta:2018vmq,Banihashemi:2018oxo,Banihashemi:2018has,Farhang:2020sij,Banihashemi:2020wtb,Visinelli:2019qqu,Akarsu:2019hmw,Koksbang:2019glb,Ye:2020btb,Ye:2020oix,Perez:2020cwa,Calderon:2020hoc,Paliathanasis:2020sfe,Bonilla:2020wbn,Vazquez:2012ag,Akarsu:2020yqa,LinaresCedeno:2021aqk,Zhou:2021xov,LinaresCedeno:2020uxx,DiValentino:2020naf,Akarsu:2021fol,Ozulker:2022slu,DiGennaro:2022ykp,Akarsu:2022typ,Colgain:2022rxy,Malekjani:2023dky,Escamilla:2023shf}, and/or an oscillatory EoS parameter that can even cross the phantom divide line (PDL)~\cite{Zhao:2017cud,Escamilla:2021uoj,Hojjati:2009ab,Xia:2006rr,Bamba:2010zxj,Pace:2011kb,farajollahi2012cosmic,Cicoli:2018kdo,Ruchika:2020avj,Heisenberg:2022gqk,Tamayo:2019gqj,Lazkoz:2010gz,Pan:2017zoh,Matsumoto:2017qil,Adil:2021zxp,Vazquez:2023kyx,Escamilla:2023shf}, and/or an oscillatory density~\cite{Colgain:2021pmf,Pogosian:2021mcs,Raveri:2021dbu,Wang:2018fng,Koksbang:2019glb,Escamilla:2021uoj,Bernardo:2021cxi,Tamayo:2019gqj,Kazantzidis:2020xta,Escamilla:2023shf}.
In this paper, we introduce and focus on a class of DE models that can exhibit very rich phenomena by incorporating all of the above mentioned behaviors to a single expansion history; they will be dubbed \textit{omnipotent dark energy} models due to their immense phenomenological capabilities. We will demonstrate their capacity to effectively alleviate the cosmological tensions by considering the Di Valentino-Mukherjee-Sen (DMS20) DE parametrization introduced in Ref.~\cite{DiValentino:2020naf} as an example.

The structure of the paper is as follows: in \cref{sec:omnipotent}, the family of omnipotent dark energy models is motivated and introduced; in \cref{sec:DMS20}, the DMS20 parametrization and its preliminary analysis are presented; in \cref{sec:method}, the datasets used and the results obtained are described; and in \cref{sec:conc}, the conclusions we derived are presented.

\section{Omnipotent dark energy}
\label{sec:omnipotent}
 There is a consensus, based on plethora of observational results~\cite{SupernovaSearchTeam:1998fmf,SupernovaCosmologyProject:1998vns,Planck:2018vyg,eBOSS:2020yzd,DES:2021wwk,DES:2022ygi,ACT:2020gnv,SPT-3G:2021wgf}, that the present-day energy budget of the Universe is dominated by the so-called dark energy with a positive density ($\rho_{\rm DE}>0$) and an EoS parameter that is about minus unity ($w_{\rm DE}\sim-1$). Due to the remarkable accuracy of $\Lambda$CDM in explaining a variety of observations, if a DE model is to dethrone the positive cosmological constant (i.e., the DE component of $\Lambda$CDM satisfying $\rho_{\rm DE}>0$ and $w_{\rm DE}=-1$), we expect the deviations between the two to either be small, be short-lived, happen in the early universe when the DE is negligible, or be a combination of these three possibilities. With that said, these conditions are barely restrictive and allow a diverse set of alternative scenarios to remedy the shortcomings of the cosmological constant. In this section, we motivate and introduce a phenomenologically very capable class of DE models whose energy densities can transition between negative and positive values and incorporate oscillatory/nonmonotonic evolution, and whose EoS parameters can attain any value and feature singularities and PDL crossings; we dub them \textit{omnipotent dark energy} models. Of course, an omnipotent DE does not submit to the conventional bounds on its energy density or EoS parameter such as the energy conditions; hence the naming ``omnipotent" that stands for unlimited power and authority. However, even without debating the validity of these bounds/conditions, their violation is not alarming when the omnipotent DE is treated as an effective source in the Friedmann equations rather than a physical component of the energy-momentum tensor.
 
 The first major feature of omnipotent DE models that we will discuss is their capability to attain negative energy densities. DE densities that attain negative values in the past, appeared repeatedly in nonparametric observational reconstructions~\cite{Bonilla:2020wbn,Sahni:2014ooa,Aubourg:2014yra,Wang:2018fng,Poulin:2018zxs,Escamilla:2021uoj,Escamilla:2023shf}, and in parametric reconstructions and models that allow negative DE densities, this feature has proved to be very effective in addressing the major tensions ($H_0$, $M_B$, $S_8$) and anomalies [baryon acoustic oscillations (BAO) Ly-$\alpha$, age of the Universe, baryon density anomalies] of $\Lambda$CDM~\cite{Visinelli:2019qqu,Sen:2021wld,Calderon:2020hoc,Sahni:2014ooa,DiValentino:2020naf,Akarsu:2019hmw,Dutta:2018vmq,Akarsu:2021fol,Akarsu:2022typ,Akarsu:2019ygx,Acquaviva:2021jov,Akarsu:2022lhx,Vazquez:2023kyx}. It is informative to focus on the $H_0$ tension to demonstrate how such a DE is able to alleviate certain tensions. Let us first establish that we assume a spatially flat and uniform (isotropic and homogeneous) universe described by the spatially flat Robertson-Walker spacetime metric whose spatial sections scale with $a(t)$ where $t$ is cosmic time; and, we choose the scaling $a_0=1$ and define the cosmological redshift $z\equiv-1+1/a$ where the subscript ``$0$", here and henceforth, denotes the present-day value of any parameter it is attached to. Now, for a given value of the comoving sound horizon at last scattering, measurements of the CMB temperature anisotropy provide strict and almost model-independent constraints on the comoving angular diameter distance to last scattering, $D_M(z_*)=c\int_0^{z_{*}}H^{-1}(z)\dd{z}$, where $H(z)$ is the Hubble parameter, $c=1$ is the speed of light, and $z_*\sim1100$ is the redshift of last scattering. These constraints, along with CMB-based constraints on the physical matter densities, result in an $H_0$ prediction within $\Lambda$CDM~\cite{Planck:2018vyg,SPT-3G:2022hvq,ACT:2020gnv} that is significantly lower than the local measurements~\cite{Riess:2021jrx,Blakeslee:2021rqi,Pesce:2020xfe} (cf. the $H_0$ tension~\cite{Bernal:2016gxb,Verde:2019ivm,Riess:2019qba,DiValentino:2020zio,DiValentino:2021izs,Perivolaropoulos:2021jda,Abdalla:2022yfr,Kamionkowski:2022pkx}); also, they allow us to asses how the qualitative behavior of a dynamical DE model would impact the CMB-based $H_0$ prediction of the model. Let $H(z)$ describe a model whose DE density is $\rho_{\rm DE}(z)$, and $H_{\Lambda{\rm CDM}}(z)$ describe the $\Lambda$CDM model whose effective DE density originating from the cosmological constant is $\rho_{\rm CC}$. If $\Delta H\equiv H(z)-H_{\Lambda{\rm CDM}}(z)$ is negative for large redshifts, strict constraints on $D_M(z_*)$ imply $\Delta H>0$ must be satisfied at some lower redshifts so that the result of the integration does not substantially change (see Ref.~\cite{Akarsu:2022lhx} for in-depth discussions); this can be straightforwardly translated to $\Delta H_0>0$, i.e., a higher $H_0$ value more in line with local measurements. Due to the solid constraints also on the physical matter densities, the sign of $\Delta \rho_{\rm DE}\equiv \rho_{\rm DE}(z)-\rho_{\rm CC}$ is highly correlated with the sign of $\Delta H$, and a higher $H_0$ value can be achieved by means of a DE density that decreases towards the past (corresponding to $\Delta H<0$ at high redshifts). Phantom DE models as defined in Ref.~\cite{Caldwell:1999ew}, which are characterized by $\rho_{\rm DE}>0$ and $w_{\rm DE}<-1$, are the epitome of such models with a density that monotonically decreases towards the past, and can solve the $H_0$ tension\footnote{Unfortunately, they are known to be in disagreement with a combination of BAO and Type Ia Supernovae (SNIa) data, i.e., they suffer from the so called ``sound horizon problem"~\cite{Knox:2019rjx,Arendse:2019hev,DiValentino:2022fjm}.}~\cite{Vagnozzi:2018jhn,DiValentino:2019exe,DiValentino:2019dzu,Vazquez:2020ani,Banerjee:2020xcn,Heisenberg:2022gqk,Lee:2022cyh}. Due to this unambiguous and promising connection between higher $H_0$ values and the scaling of phantom DE densities with time, one may look for approaches that further this characteristic of the phantom models. From a purely phenomenological perspective on the background dynamics of the Universe, a DE density that monotonically decreases to achieve negative values in the past (unlike a phantom DE which is bound to have $\rho_{\rm DE}>0$) constitute a natural extension of the phantom models.
 Indeed, such models with negative DE densities amplify the characteristics of the background dynamics of phantom DE models and can achieve equivalently high $H_0$ values with even milder/slower density scaling compared to their phantom counterparts. Despite being a natural extension of phantom models in the above sense, models with negative densities are not very frequent in the cosmology literature. The $\rho_{\rm DE}(z\sim0)>0$ requirement of the present-day acceleration,\footnote{Strictly speaking, it is possible to have accelerated present-day expansion with $\rho_{\rm DE0}<0$ and $w_{\rm DE0}>0$ (discussed further in the main text), however, this case is rendered irrelevant by observations.} and the singular behavior (discussed below) of $w_{\rm DE}$ for a DE that crosses to negative densities, render these models elusive and easy to miss, particularly in studies relying on parametrization or reconstruction of $w_{\rm DE}(z)$.

The similarities between phantom DE and its negative density extensions only go so far; the inequality $w_{\rm DE}<-1$ that is the embodiment of phantom DE models, cannot hold everywhere for DE models that dynamically attain negative densities in the past. For the framework of this paper where only minimally interacting DE models within GR are considered,\footnote{This is not very restrictive since modifications to the physics underlying $\Lambda$CDM, including those from alternative theories of gravity, can be exactly mimicked at the background level with an effective DE source within GR that replaces the cosmological constant. Note however, in general, they can be distinguished with sufficiently precise tests beyond the background level~\cite{Joyce:2016vqv}.} the continuity equation in the redshift form reads, as usual,
\be
\dv{\rho_{\rm DE}(z)}{z} =  3\frac{1+w_{\rm DE}(z)}{1+z}\rho_{\rm DE}(z).\label{eq:contz}
\ee
It is clear from~\cref{eq:contz} that when $\rho_{\rm DE}<0$ is satisfied, a DE density that decreases towards the past, i.e., $\dv{\rho_{\rm DE}}{z}<0$, requires a quintessence-like EoS parameter, $w_{\rm DE}>-1$, in contrast with phantom DE models with a positive density that also decreases towards the past. Summary of all six potential 
scenarios with respect to the sign of $\rho_{\rm DE}$ and how $w_{\rm DE}$ compares to the PDL is presented in~\cref{tab:omni}. We name these scenarios in the last column of the table; ``p-" and ``positive-" stand for $\rho_{\rm DE}>0$; ``n-" and ``negative-" stand for $\rho_{\rm DE}<0$; ``CC" stands for $w_{\rm DE}=-1$; phantom stands for $w_{\rm DE}<-1$; and quintessence stands for $w_{\rm DE}>-1$. 

A DE whose density is positive today ($\rho_{\rm DE0}>0$) and is monotonically decreasing towards the past must show the usual p-phantom behavior at late times. However, if its density is to cross to negative values in the past, it should also show n-quintessence behavior at earlier times. This requires a discontinuous EoS parameter (even for a smoothly evolving $\rho_{\rm DE}$), because, as the DE density passes to negative values with increasing redshift, its EoS parameter needs to shift from $w_{\rm DE}<-1$ to the disjoint region $w_{\rm DE}>-1$ in a particular way. Let $z_{\rm p}$ be the redshift at which DE crosses to negative densities, $w_{\rm DE}$ diverges to negative infinity for $z\to z_{\rm p}^-$ and to positive infinity for $z\to z_{\rm p}^+$ (compare the energy densities of the blue and yellow curves in the top panel of~\cref{fig:demo} with their EoS parameters in the bottom panel, and see Ref.~\cite{Ozulker:2022slu} for a detailed discussion).
Note that, strictly speaking, $w_{\rm DE}$ crosses between $w>-1$ and $w<-1$ around $z_{\rm p}$; however, we use the term PDL crossing in its established sense in the literature so far, that is, only to refer to continuous crossings and not the above mentioned singular types. 

\begin{table}[t]
\caption{An omnipotent DE is any DE model that is capable of incorporating all six behaviors in this table into a single expansion scenario for at least one point in its parameter space. In the table, we drop the DE subscript in $\rho_{\rm DE}$ and $w_{\rm DE}$ for brevity.}
\begin{center}
\resizebox{0.48\textwidth}{!}{ 
\begin{tabular}{c| c|| c| c| c}
    \hline\hline
    \textbf{Density} & \textbf{EoS} & \textbf{Scaling in $\mathbf{\textit{z}}$} & \textbf{Scaling in $\mathbf{\textit{a}}$} & \textbf{Naming}\\
    \hline
    \multirow{3}{*}{$\rho>0$}&$w>-1$&$\dd{\rho}/\dd{z}>0$&$\dd{\rho}/\dd{a}<0$&p-quintessence\\ 
     &$w=-1$&$\dd{\rho}/\dd{z}=0$&$\dd{\rho}/\dd{a}=0$&positive-CC\\
    &$w<-1$&$\dd{\rho}/\dd{z}<0$&$\dd{\rho}/\dd{a}>0$&p-phantom\\
    
    \hline
    \multirow{3}{*}{$\rho<0$}&$w>-1$&$\dd{\rho}/\dd{z}<0$&$\dd{\rho}/\dd{a}>0$&n-quintessence\\
    &$w=-1$&$\dd{\rho}/\dd{z}=0$&$\dd{\rho}/\dd{a}=0$&negative-CC\\
    &$w<-1$& $\dd{\rho}/\dd{z}>0$&$\dd{\rho}/\dd{a}<0$& n-phantom\\
    \hline\hline
\end{tabular}
}
\end{center}
\label{tab:omni}
\end{table}

The second major feature of omnipotent DE models that we discuss, other than being able to attain negative energy densities, is the ability to incorporate a nonmonotonic and even oscillatory energy density evolution. Such behaviors are suggested by parametric and nonparametric reconstructions of both $\rho_{\rm DE}$ and $w_{\rm DE}$~\cite{Wang:2018fng,Tamayo:2019gqj,Escamilla:2021uoj,Colgain:2021pmf,Raveri:2021dbu,Pogosian:2021mcs,Bernardo:2021cxi,Escamilla:2023shf}; also, they naturally arise when $D_M(z)$ to a redshift is strictly constrained and these behaviors can be favored by the latest BAO measurements~\cite{Akarsu:2022lhx}. Oscillatory or just nonmonotonic evolution of a DE density that is also continuous in time (for discontinuous examples, see, e.g., Refs.~\cite{Akarsu:2021fol,Akarsu:2022typ,Ahmed:2002mj,Zwane:2017xbg,Banihashemi:2018oxo}), implies the existence of a redshift, $z_{\rm r}$, around which the scaling of the DE density with the expansion is reversed. The continuity of the density requires that the sign of $\rho_{\rm DE}$ does not change around $z_{\rm r}$, i.e., the reversal corresponds either to a transition between p-quintessence and p-phantom, or n-quintessence and n-phantom. Moreover, if the DE density is also differentiable at $z_{\rm r}$, $\eval{\dd{\rho_{\rm DE}}/\dd{z}}_{{z=z_{\rm r}}}=0$ should be satisfied, which, if $\rho_{\rm DE}$ does not vanish at $z_{\rm r}$, implies $w_{\rm DE}(z_{\rm r})=-1$ corresponding to negative-CC for a n-quintessence to n-phantom transition, and to positive-CC for a p-quintessence to p-phantom transition. In either case, under the above mentioned assumptions related to differentiability, existence of $z_{\rm r}$ is necessary for an oscillating DE density, and it is a redshift of PDL crossing.

We are now in a position to precisely define omnipotent DE models. Motivated by the above discussions on the phenomenology of DE based on observations, we define a class of DE models that can simultaneously capture in its parameter space both, transitions between negative and positive energy densities, and nonmonotonic behaviors in its evolution. \textit{An omnipotent DE is any DE model that is capable of incorporating all six combinations of $\rho_{\rm DE}<0$ and $\rho_{\rm DE}>0$ with $w_{\rm DE}<-1$, $w_{\rm DE}=-1$, and $w_{\rm DE}>-1$ into a single expansion scenario for at least one point in its parameter space.} These six combinations are summarized in~\cref{tab:omni}.

Finally, we briefly discuss the acceleration of an expanding ($\dot{a}>0$) universe in relevance with omnipotent DE. The Friedmann equations in the presence of only dust ($\rho_{\rm m}>0$ with $w_{\rm m}=0$) and DE read ${3H^2=\rho_{\rm m}+\rho_{\rm DE}}$ and ${\frac{\ddot{a}}{a}=-\frac{1}{6}\qty[\rho_{\rm m}+(1+3 w_{\rm DE})\rho_{\rm DE}]}$, where we use $8\pi G=1$ and a dot denotes ${\rm d}/{\rm d}t$. From the first Friedmann equation, a dynamical universe ($\dot{a}\neq0$) requires that ${\rho_{\rm m}>-\rho_{\rm DE}}$. From the contribution of the term relevant to the DE in the second Friedmann equation, we see that, p-quintessence with $w_{\rm DE}<-1/3$ or any p-phantom contributes positively to the acceleration as familiar to all, but, there is an extra scenario within our framework extended to negative DE densities, that is, n-quintessence with $w_{\rm DE}>-1/3$ also contribute positively to acceleration. However, recalling that $\rho_{\rm m}>-\rho_{\rm DE}$, while n-quintessence can contribute positively to acceleration with $w_{\rm DE}>-1/3$, it can result in $\ddot{a}>0$ only if ${1<-\frac{\rho_{\rm m}}{\rho_{\rm DE}}<1+3w_{\rm DE}}$ which requires $w_{\rm DE}>0$. It is important to note that this last scenario is not relevant to the present-day accelerated expansion\footnote{Still, an acceleration sourced by a negative DE density at earlier times might be worthy of consideration.} since it would imply $\rho_{\rm c0}<\rho_{\rm m0}$ (where $\rho_{\rm c0}=3H_0^2$ is the present-day critical energy density of the Universe) and result in an extremely small $H_0$ value, also it goes against the $w_{\rm DE}\sim-1$ consensus (although the validity of this assumption is dubious for negative energy densities).
Thus, it is essential that if the DE density attains negative values in the past, it should also transit to the positive regime to drive the present-day accelerated expansion. In the next section we investigate a particular omnipotent DE model previously introduced by Di Valentino, Mukherjee, and Sen in Ref.~\cite{DiValentino:2020naf}, and dubbed phantom crossing or DMS20.

\section{DMS20 parametrization}
\label{sec:DMS20}

In this section we first give a mathematical construction of an omnipotent DE model, namely, the DMS20 model, as it is introduced in Ref.~\cite{DiValentino:2020naf}, and then explore its dynamical properties in detail. DMS20 parametrizes $\rho_{\rm DE}$, such that it ensures an extremum at scale $a_m$ that satifies $\eval{\dd{\rho_{\rm DE}}/\dd{a}}_{a=a_m}=0$, and it proposes the following expression for the DE density:
\be
\begin{split}
\rho_{\rm DE}(a)=\rho_{\rm DE0}\frac{1+\alpha (a-a_m)^2 +\beta (a-a_m)^3}{1+\alpha (1-a_m)^2 +\beta (1-a_m)^3},
\label{eq:rhoExp}
\end{split}
\ee
where $\alpha$ and $\beta$ are constants associated with the polynomials of degree two and three---see Ref.~\cite{DiValentino:2020naf} for details. Notice the absence of polynomial of degree one as it vanishes (the associated coefficient is null).  It is important to note the physical significance of $a_m$. The continuity equation, $ {\dv{\rho_{\rm DE}}{a}=-\frac{3}{a}(1+w_{\rm DE})\rho_{\rm DE}}$, implies $w_{\rm DE}(a_m)=-1$ provided that $\rho_{\rm DE}$ does not vanish at the extremum $a=a_m$. Generically, $a_m$ corresponds to a crossing of the PDL; and particularly for $\alpha>0$, it is a crossing from $w_{\rm DE}>-1$ to $w_{\rm DE}<-1$ as the Universe expands (for $\alpha<0$, it crosses in the opposite direction). More generally, the EoS parameter corresponding to the DE described in~\cref{eq:rhoExp} reads, by virtue of the continuity equation,
\be
\label{eq:eos}
w_{\rm DE}(a)=-1-\frac{a[2\alpha(a-a_m)+3\beta(a-a_m)^2]}{3[1+\alpha(a-a_m)^2+\beta(a-a_m)^3]},
\ee
which yields ${w_{\rm DE}(a=0)=-1}$ and ${w_{\rm DE}({a\to\infty})=-2}$.

\begin{figure}[t]
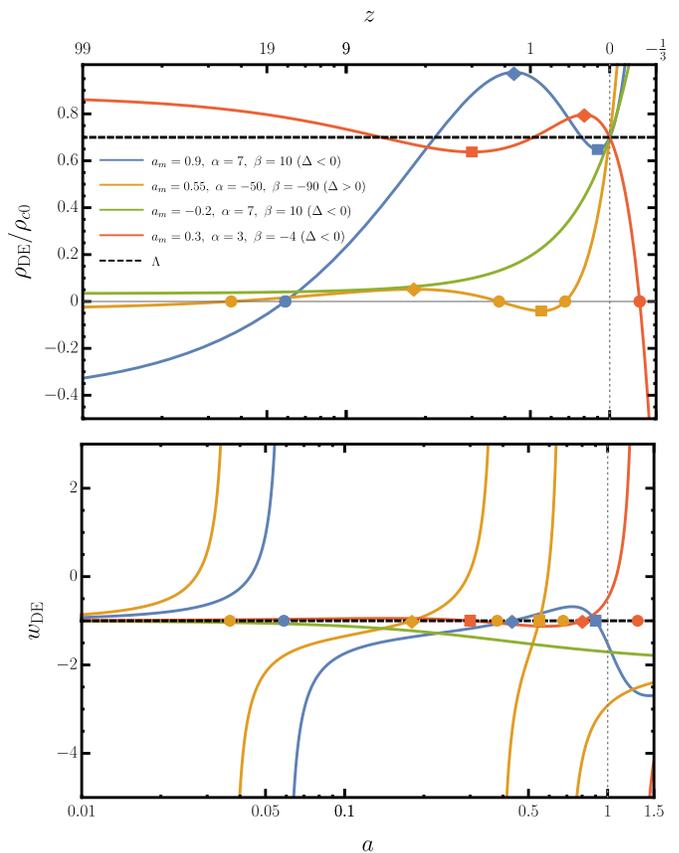

    \centering
    \makebox[0.47\textwidth][r]{\includegraphics[trim={0 0.4cm 0.1cm 0},clip,width=0.49\textwidth]{rholog_rev2.pdf}}
    \makebox[0.47\textwidth][r]{\includegraphics[width=0.48\textwidth]{eoslog_rev2.pdf}}
     \caption{The evolution of DMS20 DE densities divided by $\rho_{\rm c0}$ for various values of the free parameters $a_m$, $\alpha$ and $\beta$; and, their corresponding EoS parameters. The filled disks correspond to solutions of $a_{\rm p}$, squares to $a_m$, and diamonds to $a_n$ of the plots with matching colors. The dashed line in the top panel shows a constant energy density, hence, its EoS parameter describes the PDL. The vertical dotted line marks the present day, and on its right, we briefly extend the plots to the future. For all the plots in the figure, we use $\rho_{\rm DE0}/\rho_{\rm c0}=0.7$.}
    \label{fig:demo}
\end{figure}

The DMS20 DE introduces three extra free parameters on top of $\Lambda$, say, $\{a_m,\alpha,\beta\}$; and, some features of omnipotent DE may remain dormant for the DMS20 DE depending on the values of these parameters in two different ways. First, it may be the case that some features are completely absent; a trivial example is $\alpha=\beta=0$ in which case the DE reduces to $\Lambda$. Second, it may happen that these features are mathematically present in~\cref{eq:rhoExp,eq:eos} but they are physically less relevant since they happen at $a>1$, i.e., in the future, or they are physically irrelevant since they ``happen" at $a<0$, i.e., ``before the big bang"; two examples are $a_m>1$ and $a_m<0$ which may correspond to, respectively, a PDL crossing that has not happened yet and a virtual PDL crossing that never happens.~\cref{fig:demo} presents a nonexhaustive collection of qualitative scenarios sketching the potential behaviors the DMS20 DE can describe.

Let us be more precise with the characteristics of the DMS20 DE. The DE density in~\cref{eq:rhoExp}, being a cubic function, has at least one and at most three distinct real solutions to $\rho_{\rm DE}(a)=0$---it is possible that no real solutions exist when $\beta=0$ in which case the density is no more a cubic function. 
While it is possible that the DE density tangentially touches zero from one side at these roots for certain values of the parameters, generically  they correspond to crossings between negative and positive energy densities; see the yellow plot in the top panel of~\cref{fig:demo} for which three distinct crossings exist in the interval $0<a<1$. 
The number of real roots depends on the sign of the discriminant, ${\Delta\equiv-4\alpha^3-27\beta^2}$; if it is positive, there are three distinct real roots; if it is negative, there is only one real root; and if it is zero, there are two distinct real roots one of which has multiplicity of two and corresponds to an extremum point at which the DE density touches zero tangentially. Notice that, for $\alpha>0$, there cannot be more than one real root. One real solution, $a_{\rm p}$, to $\rho_{\rm DE}(a)=0$ is given by

\be
\resizebox{0.907\hsize}{!}{$
\begin{aligned}
a_{\rm p}=&a_m - \frac{1}{3\beta} \Bigg[ \qty( \alpha^3 + \frac{27}{2} \beta^2 - 
         \frac{3}{2} \sqrt{12 \alpha^3 \beta^2 + 81 \beta^4})^{1/3} \\
  &+ \alpha + \frac{
        \alpha^2}{\qty( \alpha^3 + \frac{27}{2} \beta^2 - 
         \frac{3}{2} \sqrt{12 \alpha^3 \beta^2 + 81 \beta^4})^{1/3}}\Bigg],
\end{aligned}
$}\label{eq:ap}
\ee
and, other real solutions, if they exist, can be found by multiplying both cubic roots in the right hand side of~\cref{eq:ap} with $e^{i2\pi/3}$ or $e^{i4\pi/3}$.
Comparing the numerator in~\cref{eq:rhoExp} with the denominator in~\cref{eq:eos}, one can observe that $w_{\rm DE}$ is singular at the points $\rho_{\rm DE}(a)=0$ is satisfied, whether the DE density crosses or tangentially touches zero; see, e.g., \cref{fig:demo} where the roots are shown with filled disks.
These singularities, are exactly of the same kind with those mentioned in~\cref{sec:omnipotent} 
and has the characteristic property of diverging to opposite infinities around the points $\rho_{\rm DE}$ vanishes, i.e., ${\lim_{a\to a_{\rm p}^\pm}w_{\rm DE}(a)=\mp\infty}$.

Further focusing on the EoS parameter given in~\cref{eq:eos}, it can easily be shown that $w_{\rm DE}(a)=-1$ admits up to three distinct solutions, namely, $a=a_m$, $a=a_m-2\alpha/3\beta$, and $a=0$. From this, we define the second scale of PDL crossing 
\be
a_n\equiv a_m-\frac{2\alpha}{3\beta},
\ee
complementary to the previously defined $a_m$---for the examples in \cref{fig:demo}, $a_m$ are indicated with filled squares, and $a_n$ with filled diamonds. While both $a_m$ and $a_n$ are PDL crossings, the direction of their crossings depend on $\alpha$ and $\beta$. For  $a_m$, the direction depends only on ${\rm sgn}(\alpha)\equiv\abs{\alpha}/\alpha$; as the scale factor passes $a_m$ with the expansion, $w_{\rm DE}$ crosses from the region $(w_{\rm DE}+1){\rm sgn}(\alpha)<0$ to $(w_{\rm DE}+1){\rm sgn}(\alpha)>0$. When $\alpha>0$, i.e. ${\rm sgn}(\alpha)=1$, the direction of crossing for $a_n$ is opposite to that of $a_m$, i.e., the EoS parameter crosses from $w_{\rm DE}<-1$ to $w_{\rm DE}>-1$ around $a_n$ as the Universe expands. On the other hand, when $\alpha<0$, the direction of crossing for $a_n$ depends also on $\beta^2$; if $27\beta^2/4<-\alpha^3$, i.e., $\Delta>0$, the crossing is now from $w_{\rm DE}>-1$ to $w_{\rm DE}<-1$, otherwise it is the same as the $\alpha>0$ case.\footnote{All of the described directions in this paragraph are correct only when these PDL crossings exist, i.e., when their corresponding scale factors are greater than zero.} Note that the sign of $\alpha/\beta$ determines which one of $a_m$ and $a_n$ is greater, hence the sign of $\beta$ is also of great importance to determine the qualitative behavior of the DMS20 DE model. We also mention the two special cases: $\alpha=0$, for which $a_m$ and $a_n$ coincide as the inflection point of the DE density with no crossing of the PDL, and $\beta=0$, for which $a_n$ does not exist. These behaviors are summarized in \cref{tab:dms20}, where, for different regions of the parameter space of $\{\alpha,\beta\}$, we show the number of real roots of $a_{\rm p}$, the direction of the PDL crossing for $a_m$ and $a_n$, and also the order of these PDL crossings. The blue and green plots in \cref{fig:demo} correspond to the first row of the table, the yellow plot corresponds to the last row, and the orange plot corresponds to the second row.

\begin{table}[t]
\caption{Summary of different qualitative behaviors of the DMS20 model for different regions of its $\{\alpha,\beta\}$ parameter space. We show the number of real roots for $\rho_{\rm DE}(a)=0$ with the $\#a_{\rm p}$ column. \protect\pdlup{} denotes a PDL crossing from $w_{\rm DE}<-1$ to $w_{\rm DE}>-1$ as the Universe expands and \protect\pdld{} denotes the other direction; the respective columns show the scales that correspond to the associated direction of PDL crossing. The ``order" column indicates which of the scale factors $a_m$ and $a_n$ is greater. Empty cells indicate that these cases are not applicable.}
\begin{center}
\resizebox{0.48\textwidth}{!}{ 
\begin{tabular}{c| c| c||c|c| c| c}
    \hline\hline
     \textbf{Discriminant} & $\bm{\alpha}$ &$\bm{\beta}$&  $\bm{\#a_{\rm p}}$ & \pdlupbf{} & \pdldbf{}&\textbf{Order}\\
    % Discriminant & $\alpha$ &$\beta$&  $\#a_{\rm p}$ & \pdlup{} & \pdld{} & Order \\
    \hline
    \multirow{4}{*}{$\frac{27\beta^2}{4}>-\alpha^3\,\,(\Delta<0)$}&\multirow{2}{*}{$\alpha>0$}&$\beta>0$&1&$a_m$&$a_n$&$a_m>a_n$\\ 
    &&$\beta<0$&1&$a_m$&$a_n$&$a_n>a_m$\\
    &\multirow{2}{*}{$\alpha<0$}&$\beta>0$&1&$a_n$&$a_m$&$a_n>a_m$\\
    &&$\beta<0$&1&$a_n$&$a_m$&$a_m>a_n$\\
    
    \hline
\multirow{4}{*}{$\frac{27\beta^2}{4}<-\alpha^3\,\,(\Delta>0)$}&\multirow{2}{*}{$\alpha>0$}&$\beta>0$&&&&\\ 
    &&$\beta<0$&&&&\\
    &\multirow{2}{*}{$\alpha<0$}&$\beta>0$&3&&$a_m$, $a_n$&$a_n>a_m$\\
    &&$\beta<0$&
    3&&$a_m$, $a_n$&$a_m>a_n$\\
    
    \hline\hline
\end{tabular}
}
\end{center}
\label{tab:dms20}
\end{table}

The last root satisfying $w_{\rm DE}(a)=-1$ is ${a=0}$, i.e., for early times ($a\rightarrow 0$, $z\to\infty$), the DE EoS parameter approaches to that of the cosmological constant\footnote{Note that, even though DMS20 DE satisfies $w_{\rm DE}(a=0)=-1$, the derivative $\eval{\dv{\rho_{\rm DE}}{a}}_{a=0}=\frac{\rho_{\rm DE0}(3\beta a_m^2-2\alpha a_m)}{1+\alpha (1-a_m)^2 +\beta (1-a_m)^3}$ does not vanish (unlike it would for the cosmological constant) except for special values of the free parameters. Nevertheless, the corresponding derivative $\eval{\dd{\rho_{\rm DE}}/\dd{z}}_{{z\to\infty}}=0$ still vanishes by virtue of the scaling between $a$ and $z$ which results in $\eval{\dd{a}/\dd{z}}_{z\to\infty}=0$. } (${w_{\rm DE}\to-1}$). The EoS parameter should satisfy this property for any DE density that attains a finite nonvanishing value as ${a\to0}$, an example of which is of course DMS20; otherwise, the integral in the right-hand side of  ${\abs{\rho_{\rm DE}(a=0)}=\abs{\rho_{\rm DE}(a_i)}{\rm 
exp}\qty[3\int_{0}^{a_i}\dd{a}\frac{1+w_{\rm DE}(a)}{a}]}$ would not converge for a typical scale $a_i$---this equation can be obtained by integrating the continuity equation. This last point is interesting because a parametrization of $\rho_{\rm DE}$ with the scale factor would generically lead to a finite and nonzero $\rho_{\rm DE}(a=0)$ implying $w_{\rm DE}(a=0)=-1$; thus, a broad class of models need to exhibit cosmological constant-like behavior in the far past.

\subsection{Decomposition into negative cosmological constant and a non--negative dynamical DE}
\label{subsec:decomposition}
Any $\rho_{\rm DE}(a)$ can phenomenologically be decomposed into an arbitrarily valued $\Lambda$ (with an effective energy density $\rho_\Lambda$) and a new DE source whose density is $\rho_{\rm DE}(a)-\rho_\Lambda$.
Such a decomposition is particularly meaningful in the presence of negative DE densities; because, while a physical fluid with a negative energy density is eccentric to say the least, anti-de Sitter (AdS) vacua (provided by $\Lambda<0$) is welcome due to the celebrated AdS/CFT correspondence  and is preferred by string theory and string-theory-motivated supergravities, and a geometric $\Lambda$ term with a negative value in the EFE or a quantum field whose potential has a negative minimum are completely viable~\cite{Maldacena:1997re,Bousso:2000xa}. For works that consider the possibility of the existence of a DE with a positive definite energy density combined with a negative cosmological constant, we refer readers to Refs.~\cite{Sen:2021wld,Dutta:2018vmq,Visinelli:2019qqu,Calderon:2020hoc,Adil:2023ara} (see also Refs.~\cite{Akarsu:2022typ,Akarsu:2021fol,Akarsu:2019hmw,Aubourg:2014yra}).

The DMS20 DE density is bounded from below for ${a\in[0,1]}$; thus, it is always possible to decompose it as described above into a $\Lambda$ whose effective energy density, $\rho_\Lambda$, corresponds to the minimum of $\rho_{\rm DE}$ (not necessarily negative) for ${a\in[0,1]}$, and a non-negative dynamical DE denoted DE$^+$, i.e., $\rho_{\rm DE}=\rho_\Lambda+\rho_{\rm DE^+}(a)$. Thus, if the minimum of $\rho_{\rm DE}(a\in[0,1])$ is at $a=a_{\rm min}$, we have $\rho_{\rm DE}(a_{\rm min})=\rho_\Lambda$ and $\rho_{\rm DE+}(a_{\rm min})=0$. Pleasantly, the minimum of $\rho_{\rm DE}(a\in[0,1])$ happens to be $\rho_{\rm DE}(a=0)$ (i.e., $a_{\rm min}=0$) for a large section of its observationally allowed parameter space based on the analyses in Ref.~\cite{DiValentino:2020naf}. For this special case, the effective energy density of the cosmological constant reads
\be
\rho_\Lambda=\rho_{\rm DE0}\frac{1+\alpha a_m^2 -\beta a_m^3}{1+\alpha (1-a_m)^2 +\beta (1-a_m)^3},
\ee
and the newly defined DE$^+$ density, $\rho_{\rm DE^+}\equiv \rho_{\rm DE}-\rho_\Lambda$ reads
\be
\resizebox{0.907\hsize}{!}{$\rho_{\rm DE^+}=\rho_{\rm DE0}\frac{\alpha (a^2-2 a a_m) +\beta (a^3-3a^2 a_m+3a a_m^2)}{1+\alpha (1-a_m)^2 +\beta (1-a_m)^3},$}
\ee 
which corresponds to the EoS parameter
\be 
\resizebox{0.907\hsize}{!}{$w_{\rm DE^+}(a)=-1-\frac{2\alpha (a-  a_m) +3\beta (a- a_m)^2}{3[\alpha (a-2  a_m) +\beta (a^2-3a a_m+3 a_m^2)]},$}
\ee 
that yields $w_{\rm DE^+}(a\to\infty)=-2$ just like $w_{\rm DE}$ itself but with $w_{\rm DE^+}(a=0)=-4/3$ instead, cf.~\cref{eq:eos}. Notice that, in this framework, $a_m$ and $a_n$ are solutions to $w_{\rm DE^+}(a)=-1$, i.e., they still correspond to scale factors of PDL crossing for the non-negative DE in this decomposition approach.

\section{Observational constraints}
\label{sec:method}

The Friedmann equation of the model we reach by replacing $\Lambda$ of the standard cosmological model with the DMS20 DE parametrization is given by 
\begin{equation}
\frac{H^2}{H_0^2}=\Omega_{\rm m0}a^{-3}
+\Omega_{\rm r 0}a^{-4}+\Omega_{\rm DE0}f(a), 
\end{equation}
where
\be
f(a)=\frac{1+\alpha (a-a_m)^2 +\beta (a-a_m)^3}{1+\alpha(1-a_m)^2 +\beta(1- a_m)^3},
\ee
and the subscript ``m" denotes all matter (baryonic and cold dark matter), and ``r" denotes radiation (photons and other relativistic relics).  Here $\Omega_{i0}\equiv\rho_{i0}/3H_0^2$ are the present-day values of the density parameters and we work in the units $8\pi G=c=1$ with $H_0=H(z=0)$ being the Hubble constant.

\subsection{Data and Methodology}
\label{data}

In order to derive constraints on the model baseline, we use the following datasets:
\begin{itemize}
\item \textbf{CMB}: From the \textit{Planck} 2018 legacy data release, we use the CMB measurements, viz., high-$\ell$ \texttt{Plik} TT likelihood (in the multipole range $30 \leq \ell \leq 2508$), TE and EE (in the multipole range $30 \leq \ell \leq 1996$),  low-$\ell$ TT-only ($2 \leq \ell \leq 29$), the low-$\ell$ EE-only ($2 \leq \ell \leq 29$) likelihood~\cite{Planck:2019nip}, in addition to the CMB lensing power spectrum measurements \cite{Planck:2018lbu}. We refer to this dataset as \texttt{Planck}.

\item \textbf{BAO}: From the latest compilation of BAO distance and expansion rate measurements from the SDSS Collaboration, we use 14 BAO measurements, viz., the isotropic BAO measurements of $D_V(z)/r_d$ [where $D_V(z)$ and $r_d$ are the spherically averaged volume distance, and sound horizon at baryon drag, respectively] and anisotropic BAO measurements of $D_M(z)/r_d$ and $D_H(z)/r_d$ [where $D_M(z)$ and $D_H(z)=c/H(z)$ are the comoving angular diameter distance and  the Hubble distance, respectively], as compiled in Table 3 of \cite{Alam_2021}. We refer to this dataset as \texttt{BAO}.

\item \textbf{Type Ia supernovae and Cepheids}: We use the SNe Ia distance moduli measurements from the Pantheon+ sample \cite{Brout_2022}, which consists of 1701 light curves of 1550 distinct SNe Ia ranging in the redshift interval $z \in [0.001, 2.26]$. We refer to this dataset as \texttt{PantheonPlus}. We also consider the SH0ES Cepheid host distance anchors, which facilitate constraints on both $M_B$ and $H_0$. When utilizing SH0ES Cepheid host distances, the SNe Ia distance residuals are modified following the relationship Eq.~(14) of Ref.~\cite{Brout_2022}. We refer to this dataset as \texttt{PantheonPlus\&SH0ES}. 
\end{itemize}

In our analyses, we allow the parameters to follow the flat uniform priors described in~\cref{tab:priors}. We study an extended cosmological model that in addition to the six parameters of the standard $\Lambda$CDM model, i.e., the baryon and cold dark matter densities $\Omega_{\rm b}h^2$ and $\Omega_{\rm c}h^2$, the optical depth $\tau$, the amplitude and spectral index of the scalar fluctuations $\log[10^{10}A_{s}]$ and $n_s$, and the angular size of the horizon at the last scattering surface $\theta_{\rm{MC}}$, is considering the three parameters of the DMS20 DE parametrization in~\cref{eq:rhoExp}, i.e., $\alpha$, $\beta$, and $a_m$. Following the analyses in Ref.~\cite{DiValentino:2020naf}, we choose the same priors for all nine free parameters; in particular, we restrict $\alpha$ and $\beta$ to be non-negative, and assume $a_m$ is a physical scale that existed in the expansion history of the Universe. With these choice of priors, we restrict the model behavior to the first row of \cref{tab:dms20}, i.e., $a_m$ is a PDL crossing from $w_{\rm DE}>-1$ to $w_{\rm DE}<-1$ (as the Universe expands) and $a_n$ from $w_{\rm DE}<-1$ to $w_{\rm DE}>-1$; note that $a_n<a_m$ should be satisfied for these priors but $a_n$ might not exist physically (i.e, in the range [0,1]) as it is unbounded from below. Moreover, only a single real root exists for the derived parameter $a_{\rm p}$ and it is also unbounded from below---see \cref{sec:DMS20} for more details on how the choice of $\alpha$ and $\beta$ affects the physical meaning and existence of other parameters.
We ran our modified version of the publicly available \texttt{CLASS+MontePython} code \cite{Blas_2011,Audren:2012wb,Brinckmann:2018cvx} that implements the model described in~\cref{eq:rhoExp}, using Metropolis-Hastings mode to derive constraints on cosmological parameters. All of our runs reached a Gelman-Rubin convergence criterion of $R - 1 < 10^{-2}$.\\

In order to compare the goodness of the statistical fits of the DMS20 DE with respect to the standard $\Lambda$CDM scenario, we make use of Akaike information criterion (AIC) and $\log$-Bayesian evidence along with $\chi^2_{\rm min}= -2 \ln \mathcal{L}_{\rm max}$ with $\mathcal{L}$ being the likelihood, see~\cref{tab:constraints}. In particular, we first give the relative best fit (${\Delta\chi^2_{\rm min}=\chi^2_{\rm min,DMS20}-\chi^2_{{\rm min,}\Lambda{\rm CDM}}}$) and then the relative AIC (${\Delta{\rm AIC}={\rm AIC}_{\rm DMS20}-{\rm AIC}_{\Lambda{\rm CDM}}}$), where
\begin{eqnarray}
    {\rm AIC }= \chi^2_{\rm min} + 2K
\end{eqnarray}
with $\chi^2_{\rm min}$ being the term which incorporates the goodness-of-fit through the likelihood $\mathcal{L}$ and $2K$ being the term which is interpreted as the penalization factor given by two times the number of parameters ($K$) of the model. The preferred model is the one with the smaller AIC value, and therefore negative values of $\Delta{\rm AIC}$ imply support for DMS20 DE against the $\Lambda$CDM scenario, and the lower (more negative) the $\Delta{\rm AIC}$ value, the stronger support there is. 

We then investigate the relative log-Bayesian evidence used to perform model comparison via the Jeffreys' scale; we compute it using the publicly accessible MCEvidence package\footnote{\href{https://github.com/yabebalFantaye/MCEvidence}{MCEvidence Package}}(\cite{Heavens:2017hkr,Heavens:2017afc}). In general, for a dataset $D$ and a given model $\mathcal{M}_i$ with a set of parameters $\Theta$, Bayes' theorem results in
\begin{equation}
P(\Theta|D, \mathcal{M}_i) = \frac{\mathcal{L}(D|\Theta, \mathcal{M}_i) \pi(\Theta|\mathcal{M}_i)}{\mathcal{E}(D|\mathcal{M}_i)},
\end{equation}
where $P(\Theta|D, \mathcal{M}_i)$ is the posterior probability distribution function of the parameters, $\pi(\Theta|\mathcal{M}_i)$ is the prior for the parameters, $\mathcal{L}(D|\Theta, \mathcal{M}_i)$ is the likelihood function, and $\mathcal{E}(D|\mathcal{M}_i)=\int_{\mathcal{M}_i} \mathcal{L}(D|\Theta, \mathcal{M}_i) \pi(\Theta|\mathcal{M}_i) \text{d}\Theta$ is the Bayesian evidence. To make a comparison of the model $\mathcal{M}_i$ with some other model $\mathcal{M}_j$, we compute the ratio of the posterior probabilities of the models, given by
\begin{equation}
\frac{P(\mathcal{M}_i|D)}{P(\mathcal{M}_j|D)} = B_{ij}\frac{P(\mathcal{M}_i)}{P(\mathcal{M}_j)},
\end{equation}
where $B_{ij}$ is the Bayes' factor given by
\begin{equation}
    B_{ij}=\frac{\mathcal{E}(D|\mathcal{M}_i)}{\mathcal{E}(D|\mathcal{M}_j)}\equiv\frac{\mathcal{Z}_i}{\mathcal{Z}_j}.
\end{equation}
Accordingly, we define the relative log-Bayesian evidence for the DMS20 DE with respect to the $\Lambda$CDM scenario as follows:
\begin{equation}\label{eq:evidence}
    \Delta\ln\mathcal{Z}\equiv\ln B_{\rm DMS20,\Lambda CDM}=\ln \mathcal{Z}_{\rm DMS20}-\ln \mathcal{Z}_{\rm \Lambda CDM} .
\end{equation}
As the model with smaller value of $|\ln \mathcal{Z}|$ is the preferred one, like in the case of $\Delta{\rm AIC}$, the negative values of $\Delta\ln\mathcal{Z}$ imply support for DMS20 DE against the $\Lambda$CDM scenario, and the lower (more negative) the $\Delta\ln\mathcal{Z}$ value, the stronger support there is. The rule of thumb in the literature for interpreting $|\ln \mathcal{Z}|$ with reference to the revised Jeffreys' scale given in Ref.~\cite{Trotta:2008qt} is that the evidence is inconclusive if $0 \leq |\ln B_{ij}|  < 1$, weak if $1 \leq | \ln B_{ij}|  < 2.5$, moderate if $2.5 \leq |\ln B_{ij}|  < 5$, strong if $5\leq| \ln B_{ij} | < 10$, and very strong if $10\leq| \ln B_{ij} |$, in favor of the preferred model.

\begin{table}[t!]
\caption{We use the same flat priors in all the analyses.}
\begin{center}
\begin{tabular}{c|c}
Parameter                    & Prior\\
\hline 
$\Omega_{b} h^2$             & $[0.005,0.1]$\\
$\Omega_{c} h^2$             & $[0.005,0.1]$\\
$\tau$                       & $[0.01,0.8]$\\
$n_s$                        & $[0.8,1.2]$\\
$\log[10^{10}A_{s}]$         & $[1.6,3.9]$\\
$100\theta_{MC}$             & $[0.5,10]$\\ 
$\alpha$                        & $[0,30]$\\ 
$\beta$                        & $[0,30]$\\
$a_m$                        & $[0,1]$\\
\end{tabular}\qquad\qquad
\end{center}
\label{tab:priors}
\end{table}

%%%%%%%%%%%%%%%%%%%%%%%%%%%%%%%%%%
\begin{table*}[ht!]
\caption{{\small 68\% CL constraints on the cosmological parameters for the different Planck dataset combinations explored in this work, for the DMS20 DE model. 
In the last three rows, the relative best fit (${\Delta\chi^2_{\rm min}=\chi^2_{\rm min,DMS20}-\chi^2_{{\rm min,}\Lambda{\rm CDM}}}$), Akaike information criterion (${\Delta{\rm AIC}={\rm AIC}_{\rm DMS20}-{\rm AIC}_{\Lambda{\rm CDM}}}$), and $\log$-Bayesian evidence ($\Delta\ln\mathcal{Z} = \ln \mathcal{Z}_{\Lambda \rm CDM} - \ln \mathcal{Z}_{\rm DMS20}$) are also displayed; negative values are in favor of the DMS20 DE model against the standard $\Lambda$CDM scenario.}}
\begin{center}
\resizebox{0.80\textwidth}{!}{  
\begin{tabular}{ c |c c c c  c c c c } 
  \hline
 \hline
  & Planck & Planck  & Planck & Planck+BAO \\ 
  Parameters & +BAO  & +PantheonPlus & +PantheonPlus\&SH0ES & +PantheonPlus\&SH0ES \\ 
 \hline 
  $\Omega_{\rm c} h^2$  & $0.1196\pm 0.0011          $ & $0.1193\pm 0.0013          $ &$0.1176\pm 0.0011          $ & $0.1198\pm 0.0011          $ \\
   $10^2 \Omega_{\rm b} h^2$ & $2.241\pm 0.014            $ & $2.243\pm 0.015            $ &$2.257\pm 0.014            $ &$2.243\pm 0.014            $ \\
  $100\theta_{MC}$ & $1.04191\pm 0.00029        $ &$1.04193\pm 0.00029        $ & $1.04207\pm 0.00029        $ &$1.04191\pm 0.00029        $   \\
  $\tau$  & $0.0517\pm 0.0076          $ &$0.0516\pm 0.0077          $ &$0.0490^{+0.0081}_{-0.0071}$&$0.0502\pm 0.0076          $   \\
  $n_s$  & $0.9662\pm 0.0038          $ & $0.9669\pm 0.0043          $ &$0.9714\pm 0.0041          $ &$0.9659\pm 0.0038          $  \\
  ${\rm{ln}}(10^{10}A_s)$  & $3.038\pm 0.015            $ &$3.037\pm 0.015            $ & $3.027^{+0.016}_{-0.014}   $& $3.035\pm 0.015            $ \\
     $a_m$ & $0.841^{+0.053}_{-0.062}   $ &
      
     $0.952^{+0.042}_{-0.016} $ &$0.957^{+0.016}_{-0.023}   $ & $0.922^{+0.041}_{-0.035}   $    \\
  $\alpha$  & $7.2^{+4.8}_{-3.9}         $ &$3.4^{+1.5}_{-2.5}         $ & $7.0^{+1.6}_{-2.0}         $ &$<2.77       $   \\
  $\beta$  & $>11.1               $&$7.3^{+3.0}_{-6.0}         $ & $16.5^{+3.4}_{-4.3}        $ &$6.5^{+1.9}_{-3.4}         $ \\
   \hline
   $H_0\, {\rm[km/s/Mpc]}$ & $72.1^{+3.2}_{-4.5}        $&$68.9^{+1.4}_{-2.0}        $ &$73.49\pm 0.98             $ &$70.05\pm 0.64             $ \\
    $\Omega_{\rm m}$ & $0.277\pm 0.029            $ &$0.301^{+0.018}_{-0.015}   $ & $0.2610\pm 0.0077          $  &$0.2912\pm 0.0057          $ \\
   $\sigma_8$ &  $0.848^{+0.028}_{-0.037}   $ & $0.823^{+0.015}_{-0.019}   $& $0.861\pm 0.012            $ &$0.835\pm 0.010            $ \\
   $S_8$  &  $0.812\pm 0.017            $    & $0.823\pm 0.013            $&$0.803\pm 0.011            $ &$0.823\pm 0.011            $  \\
   $r_{\rm drag} \,{\rm[Mpc]}$  & $147.15\pm 0.25            $ &$147.20\pm 0.28            $ &$147.50\pm 0.26            $ &$147.07\pm 0.25            $  \\
  $t_0\,{\rm[Gyr]}$ &$13.740\pm 0.036           $ &$13.69^{+0.12}_{-0.07}    $ &$13.454\pm 0.056           $ &$13.679\pm 0.031           $ \\
 \hline
 $\Delta \chi^2_{\rm min}$ &  $-3.74$ &  $-2.80$ &  $-38.64$ &  $-11.74$\\
  $\Delta \rm AIC$ &  $2.26$ &  $3.20$ &  $-32.64$ &  $-5.74$
 \\

 $\Delta\ln\mathcal{Z}$ & $2.34$ & $6.39$ & $-10.44$ & $0.95$\\

  \hline
  \hline
\end{tabular}
}
\end{center}
\label{tab:constraints}
\end{table*}

%%%%%%%%%%%%%%%%%%%%%%%%%%%%%%%%%%
%%%%%%%%%%%%%%%%%%%%%%%%%%%%%%%%%%

\begin{figure*}[t]
   \includegraphics[width=0.9\textwidth]{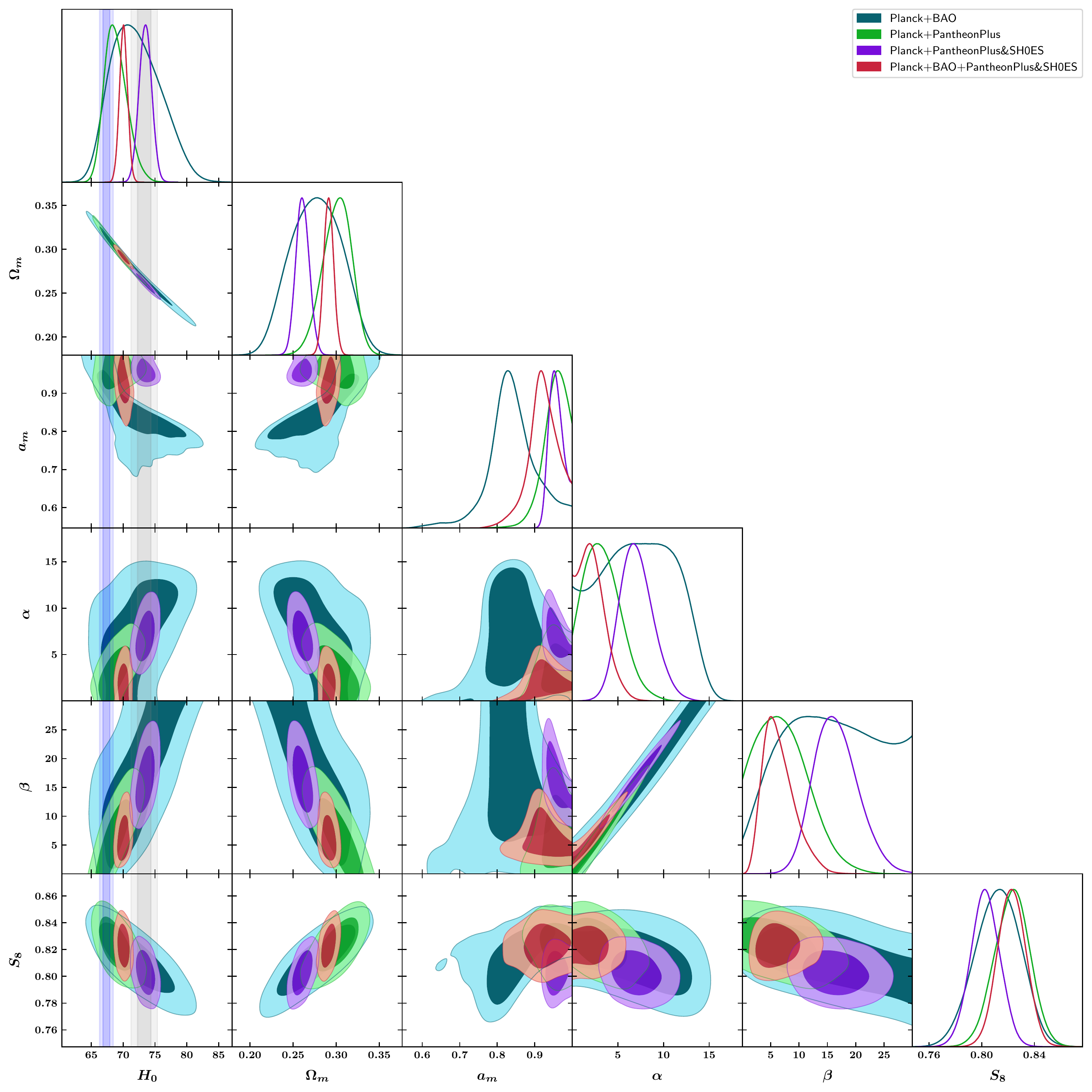}
     \caption{1D posterior distribution and 2D contour plots for a few parameters of interest.}
    \label{fig:all}
\end{figure*}

\begin{figure*}[t]
   \includegraphics[width=0.32\textwidth]{H0_ap_rev.pdf}
   \includegraphics[width=0.32\textwidth]{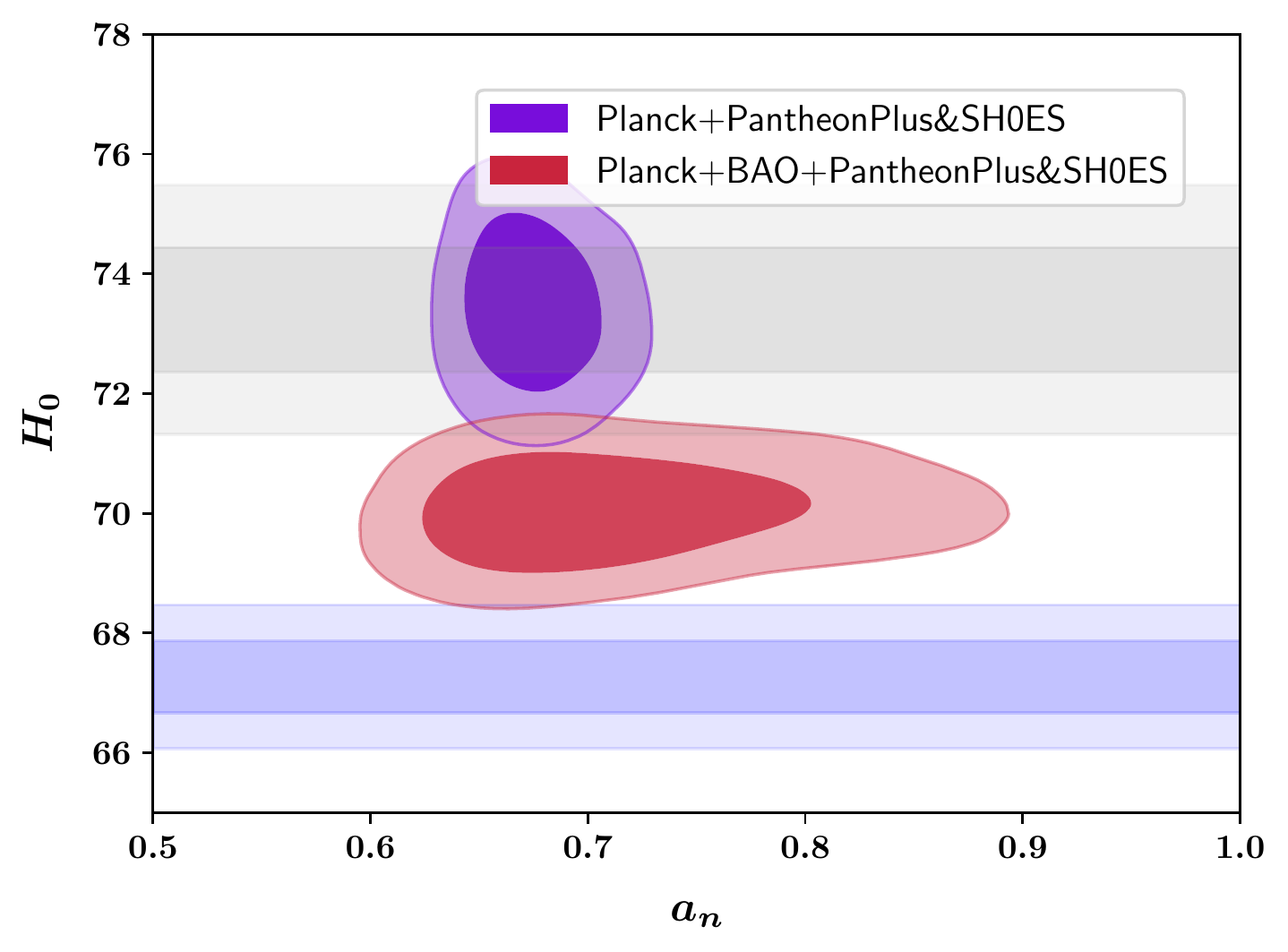}
      \includegraphics[width=0.32\textwidth]{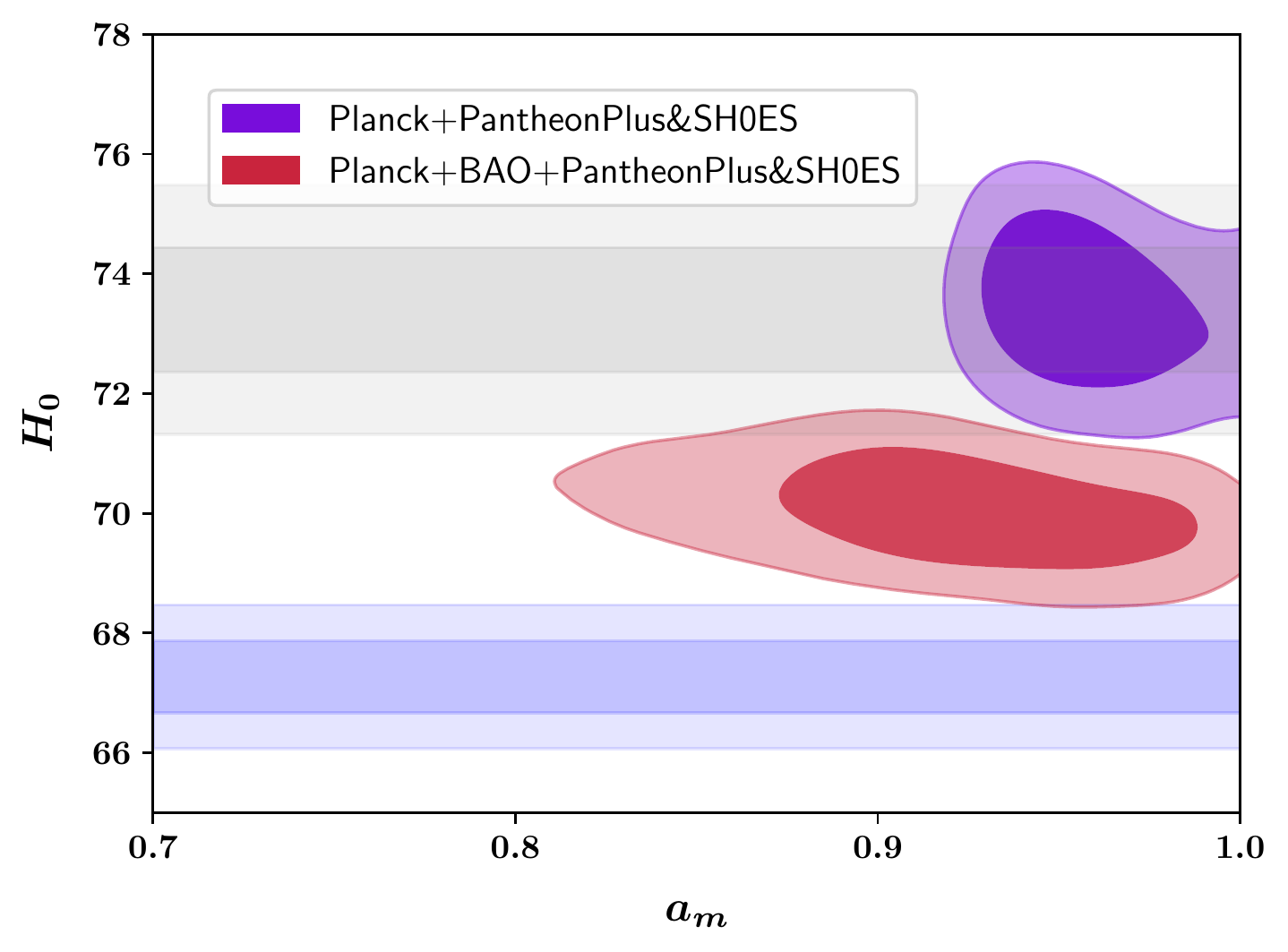}
     \caption{Two-dimensional (at 68\% and 95\% CL)  marginalized posterior distributions of $H_0$ vs. the free parameter $a_m$, and the derived parameters $a_{\rm p}$ and $a_n$. The blue band shows the $\Lambda$CDM value $H_0 = 67.27\pm 0.60{\rm \,km\, s^{-1}\, Mpc^{-1}}$ (68\% CL)~\cite{Planck:2018vyg} inferred from \textit{Planck} 2018 with $1\sigma$ and $2\sigma$ errors, and the gray band shows the SH0ES measurement $H_0 = 73.04\pm 1.04{\rm \,km\, s^{-1}\, Mpc^{-1}}$ (68\% CL)~\cite{Riess:2021jrx}.}
    \label{fig:H02D}
\end{figure*}

\begin{figure}[b]
    \centering
\includegraphics[width=0.48\textwidth]{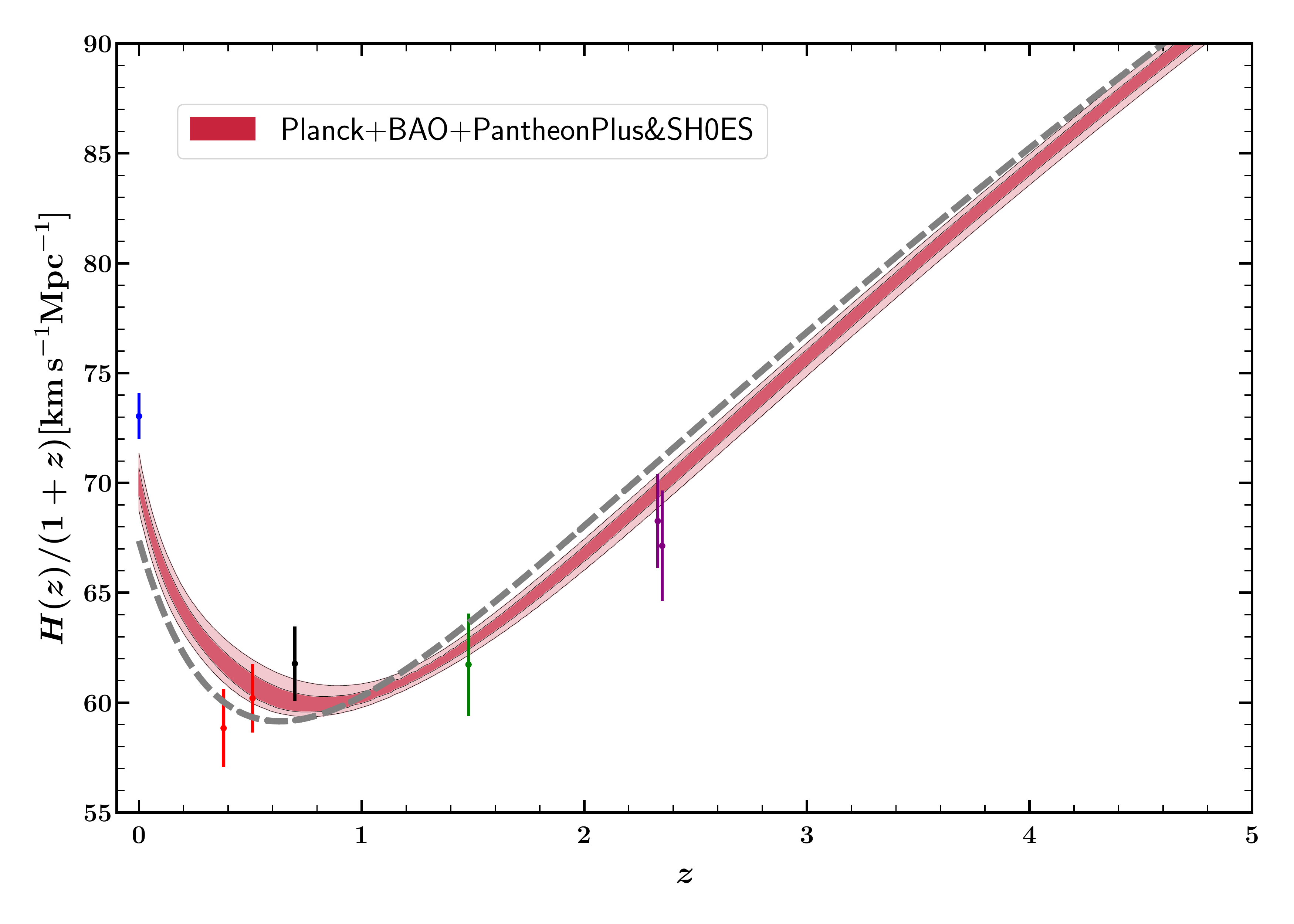}
    \caption{Posterior distributions (68\% and 95\% CL) of $H(z)/(1+z)$ for the Planck+PantheonPlus\&SH0ES (shown in red) and Planck+BAO+PantheonPlus\&SH0ES (shown in violet) data combinations. For comparison, we show the $\Lambda$CDM \texttt{Plik} best fit result from Ref.~\cite{Planck:2018vyg} with a dashed gray line. The vertical bars show various data with 68\% errors. The blue bars are the TRGB ${H_0=69.8\pm0.8\rm \,km\, s^{-1}\, Mpc^{-1}}$ measurement~\cite{Freedman:2019jwv} and the SH0ES ${H_0=73.04\pm1.04\rm \,km\, s^{-1}\, Mpc^{-1}}$ measurement~\cite{Riess:2021jrx}. The rest are BAO data that can be found in Ref.~\cite{eBOSS:2020yzd} and references therein: the red bars are the BOSS DR12 consensus Galaxy (from $z_{\rm eff}=0.38,\,0.51$) measurements; the black bar is the eBOSS DR16 LRG (from $z_{\rm eff}=0.70$) measruement; the green bar is the eBOSS DR16 Quasar (from $z_{\rm eff}=1.48$); the purple bars are the eBOSS DR16 Ly-$\alpha$-Ly-$\alpha$ (from $z_{\rm eff}=2.33$) and eBOSS DR16 Ly-$\alpha$-quasar (from $z_{\rm eff}=2.33$ but shifted to $z=2.35$ in the figures for visual clarity) measurements.}
    \label{fig:hz}
\end{figure}

% \begin{figure}[t!]
%     \centering
% \includegraphics[width=0.48\textwidth]{Hzbyz3.pdf}
%     \caption{Posterior distributions (68\% and 95\% CL) of $H(z)/(1+z)$ for the Planck+PantheonPlus\&SH0ES (shown in red) and Planck+BAO+PantheonPlus\&SH0ES (shown in violet) data combinations. For comparison, we show the $\Lambda$CDM \texttt{Plik} best fit result from Ref.~\cite{Planck:2018vyg} with a dashed gray line. The vertical bars show various data with 68\% errors. The blue bars are the TRGB ${H_0=69.8\pm0.8\rm \,km\, s^{-1}\, Mpc^{-1}}$ measurement~\cite{Freedman:2019jwv} and the SH0ES ${H_0=73.04\pm1.04\rm \,km\, s^{-1}\, Mpc^{-1}}$ measurement~\cite{Riess:2021jrx}. The rest are BAO data that can be found in Ref.~\cite{eBOSS:2020yzd} and references therein: the red bars are the BOSS DR12 consensus Galaxy (from $z_{\rm eff}=0.38,\,0.51$) measurements; the black bar is the eBOSS DR16 LRG (from $z_{\rm eff}=0.70$) measruement; the green bar is the eBOSS DR16 Quasar (from $z_{\rm eff}=1.48$); the purple bars are the eBOSS DR16 Ly-$\alpha$-Ly-$\alpha$ (from $z_{\rm eff}=2.33$) and eBOSS DR16 Ly-$\alpha$-quasar (from $z_{\rm eff}=2.33$ but shifted to $z=2.35$ in the figures for visual clarity) measurements.}
%     \label{fig:hz}
% \end{figure}

% 

\begin{figure}[t]
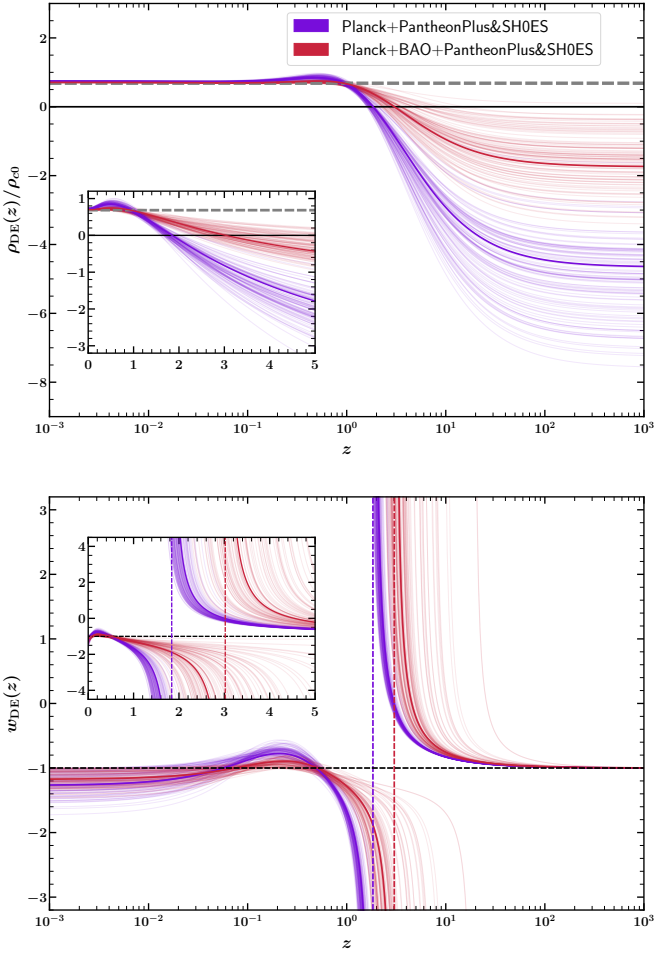

     \makebox[0.495\textwidth][r]{\includegraphics[width=0.51\textwidth]{rho_de_inset_rev.pdf}}
    \makebox[0.495\textwidth][r]{\includegraphics[width=0.51\textwidth]{w_de_inset_rev.pdf}}
     \caption{The figure shows the posteriors of $\rho_{\rm DE}(z)/\rho_{\rm c0}$ (top panels) and $w_{\rm DE}(z)$ (bottom panels); the more frequent the lines, the more probable. Violet is for the analysis with the Planck+PantheonPlus\&SH0ES dataset, and red is for Planck+BAO+PantheonPlus\&SH0ES. The bolder solid lines correspond to the best fit sample of their corresponding dataset. The horizontal gray dashed line in the top panels show the \texttt{Plik} best fit value in Ref.~\cite{Planck:2018vyg} for comparison, and the horizontal black dashed line in the bottom panels is the PDL. The vertical dashed lines in the bottom panels show the redshifts DE crosses to negative values for the best fit samples of the same color, i.e., they show the best fit $z_{\rm p}$; these dashed lines also correspond to the asymptotes of $w_{\rm DE}(z)$ for the best fit plots. The inset plots show the same posteriors but with a linear scale on the horizontal axes.}
    \label{fig:DEplots}
\end{figure}

\begin{figure}
    \centering
    \makebox[0.495\textwidth][r]{\includegraphics[trim={0 1.08cm 0 0},clip,width=0.48\textwidth]{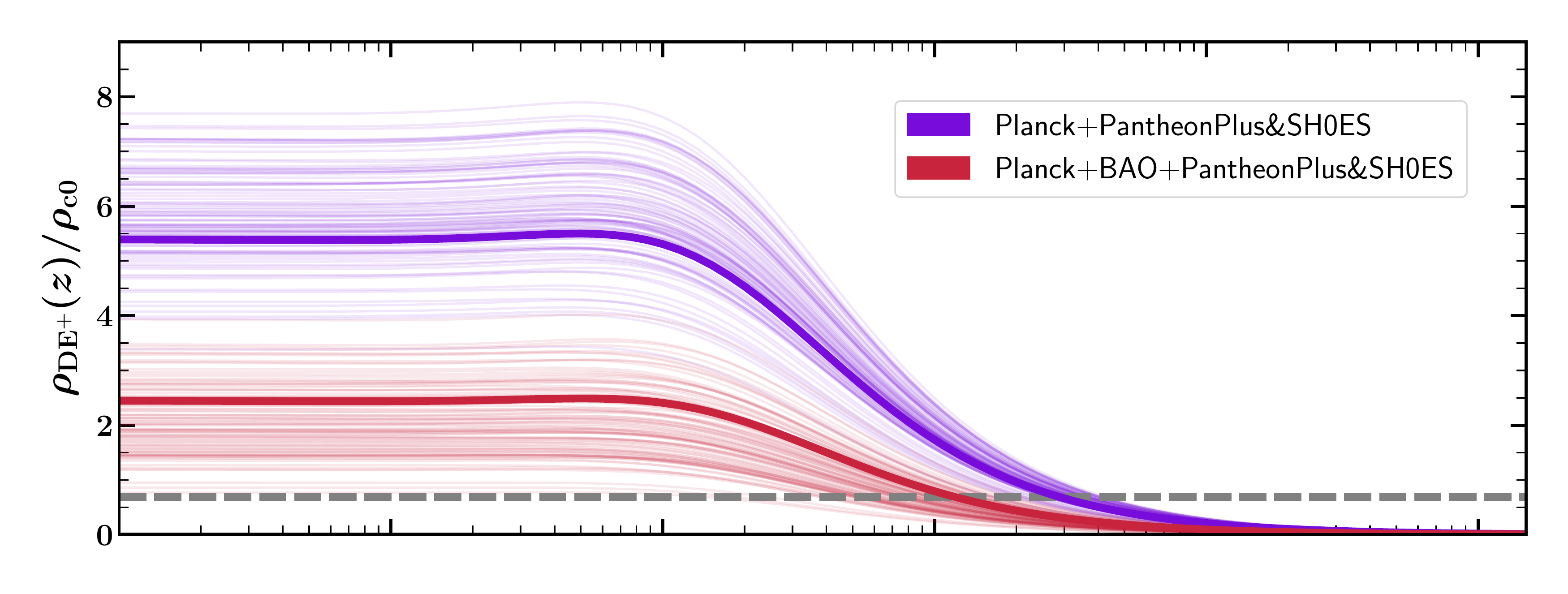}}
    \makebox[0.495\textwidth][r]{\includegraphics[width=0.49\textwidth]{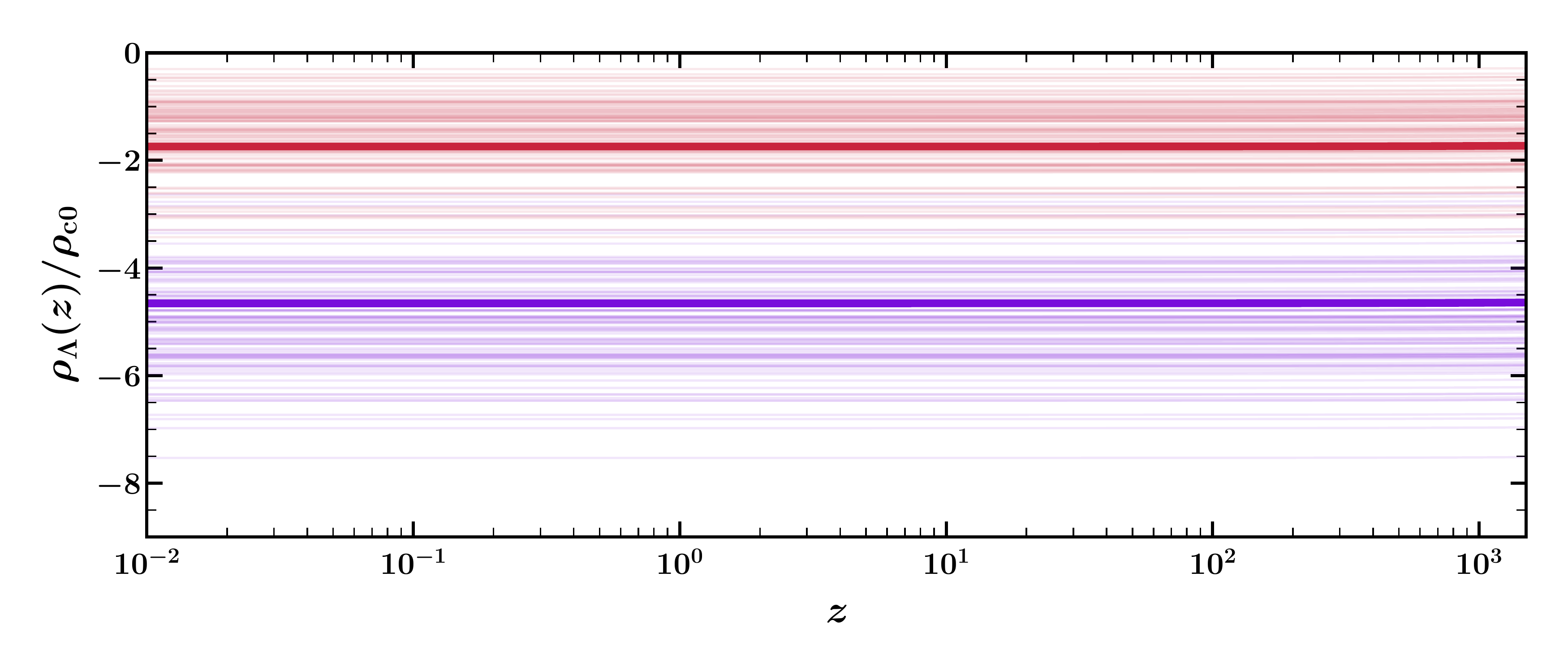}}
    \makebox[0.495\textwidth][r]{\includegraphics[trim={0 0 0 0.5cm},clip,width=0.498\textwidth]{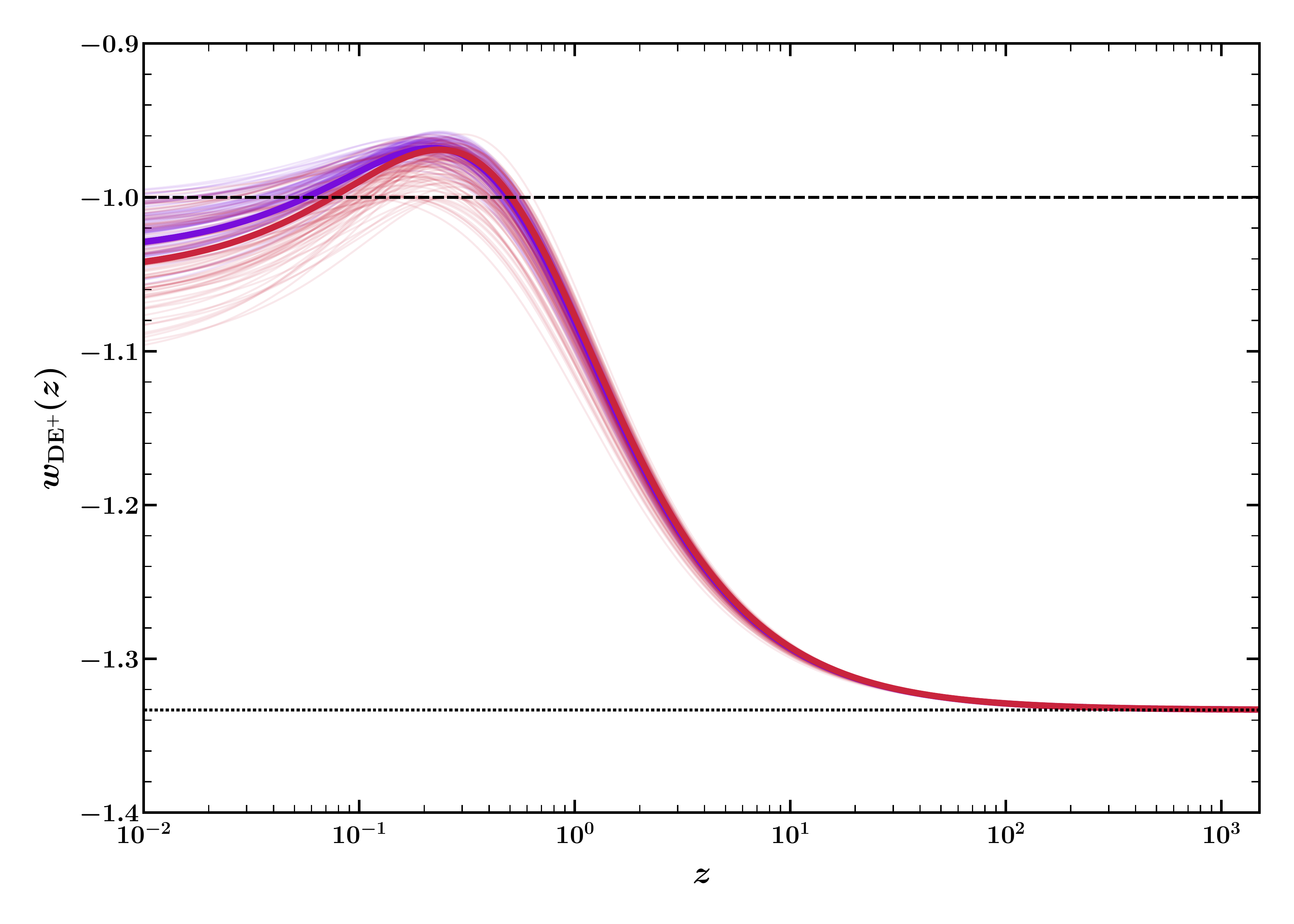}}
    
    \caption{The figure shows the posteriors of $\rho_{\rm DE^+}(z)/\rho_{\rm c0}$ and $\rho_{ \Lambda}/\rho_{\rm c0}$ on the top panels, and $w_{\rm DE^+}(z)$ on the bottom panel; the more frequent the lines, the more probable. Violet is for the analysis with the Planck+PantheonPlus\&SH0ES dataset, and red is for Planck+BAO+PantheonPlus\&SH0ES. The bolder solid lines correspond to the best fit sample of their corresponding dataset. For comparison, the horizontal gray dashed line in the top panels show the \texttt{Plik} best fit value of $\rho_\Lambda$ in Ref.~\cite{Planck:2018vyg}. In the bottom panels, the horizontal black dashed line is the PDL and the horizontal black dotted line shows the limit $w_{\rm DE^+}(a=0)=-4/3$ that holds by construction (see \cref{subsec:decomposition}).
    }\label{fig:DE+}
\end{figure}

\subsection{Results}

In this section we discuss the constraints we obtain on the DMS20 DE parametrization~\cite{DiValentino:2020naf} using a collection of recent cosmological data, as described in the previous section. We report in~\cref{tab:constraints} the constraints at 68\% confidence level (CL) for the free parameters of the model (above the horizontal line) and the derived ones (below the horizontal line) that are of interest to discuss the cosmological tensions. We remind that the DMS20 parametrization is just one example of omnipotent DE and the results obtained in this section do not apply to all omnipotent models; rather, the results hint at how the negativity and nonmonotonicity features of a DE density could interact with the data while being robust only for the DMS20 parametrization specifically.

It would be most convenient to begin our discussion on the results by making the discussion over the free parameter $a_m$, viz., the scale factor at which the usual PDL crossing (i.e., p-quintessence--p-phantom crossing) occurs; the results for this parameter are particularly important since in Ref.~\cite{DiValentino:2020naf}, the paper in which DMS20 DE parametrization was originally proposed and confronted with observational data, discussions over the results were based on this parameter. As already noticed in Ref.~\cite{DiValentino:2020naf} the Planck+BAO combination favors a crossing of the PDL in $a_m=0.841^{+0.053}_{-0.061}$ at 68\% CL and, because of their negative correlation, a value of the Hubble constant $H_0=72.1^{+3.2}_{-4.5}$ km/s/Mpc at 68\% CL, perfectly in agreement with the SH0ES measurement~\cite{Riess:2021jrx}. Moreover, thanks to the strong anticorrelation between $H_0$ and the $S_8$ parameter we can see in~\cref{fig:all}, a higher Hubble constant value corresponds to a lower $S_8$ value, helping also with the tension with the weak-lensing experiments. This result with updated and extended BAO data significantly improves the findings of Ref.~\cite{DiValentino:2020naf}. However, for this dataset combination, the DMS20 DE parametrization is moderately disfavored with respect to the $\Lambda$CDM scenario by the Bayesian evidence, because it is penalized by the Occam's razor principle with three additional parameters. The fact that some of the model parameters are unconstrained as evident from Table IV, also affects the Bayesian evidence of the DMS20 model.

On the other hand, when Planck is combined with PantheonPlus we see instead that $a_m=0.952^{+0.042}_{-0.016}$ at 68\% CL, and given that they are not correlated anymore, $H_0$ shifts down to $H_0=68.9^{+1.4}_{-2.0}$ km/s/Mpc at 68\% CL ($H_0=68.9^{+3.6}_{-3.2}$ at 95\% CL), at about $2\sigma$ tension with the SH0ES value. 
This result increases the tension with respect to the one obtained in Ref.~\cite{DiValentino:2020naf} for the previous Pantheon release, but given the moderate agreement with the local measurement, we can still safely add the SH0ES calibration and evaluate the effects on the model. Also for this dataset combination, the DMS20 DE parametrization is disfavored by the model comparison.

In the third column of~\cref{tab:constraints}, we can see that the combination Planck+PantheonPlus\&SH0ES, by fixing the $H_0$ value in agreement with the local measurements, strengthen the $1\sigma$ indication of the previous case, now giving $a_m=0.957^{+0.016}_{-0.023}$ at 68\% CL, i.e., different from $1$ at more than 1$\sigma$, but in agreement with $1$ within 2$\sigma$ ($a_m>0.929$ at 95\% CL). In this case the $S_8$ parameter goes significantly down, possibly reducing also the $S_8$ tension. Also, for this data combination resulting in an almost perfect agreement with the SH0ES $H_0$ value better than all our other analyses, the constraints on $t_0$ as well are in excellent agreement with the estimation from GCs, $t_{\rm 0}=13.50\pm0.27$~\cite{Valcin:2021jcg}, again better than rest of our data combinations---although, note that none of our analyses yield a significant discrepancy compared to this $t_0$ estimation. For this dataset combination, the DMS20 DE parametrization is instead very strongly favored with respect to the $\Lambda$CDM scenario by the Bayesian evidence, with a value $\ln B_{ij}=-10.44$.

However, the most interesting result is given by the full combination Planck+BAO+PantheonPlus\&SH0ES, reported in the last column of~\cref{tab:constraints}. In this case, contrarily to what expected, the final contours, such as $S_8$ and $H_0$ in~\cref{fig:all}, do not resemble the combination of Planck+PantheonPlus\&SH0ES and Planck+BAO, but rather the combination of Planck+PantheonPlus and Planck+BAO. This can be explained because Planck+BAO and Planck+PantheonPlus have different directions of correlation in the parameter space, and once we combine them together the final result will coincide with their combination.
In other words, the SH0ES prior is not powerful enough to shift this result more in agreement with it, and we find $H_0=70.05\pm0.64$ km/s/Mpc at 68\% CL for Planck+BAO+PantheonPlus\&SH0ES.
This can also clarify why some parameters have larger errors than the same combination without BAO data, such as $a_m=0.922^{+0.041}_{-0.035}$ at 68\% CL, because of the SH0ES prior attempting to pull the Hubble constant in its direction. For this full data combination, we also show the constraints on the Hubble function of DMS20 in~\cref{fig:hz} and the Planck best fit result from Ref.~\cite{Planck:2018vyg} for comparison. We see that, higher values of $H(z\lesssim1)$ compared to the Planck best fit are unsurprisingly compensated by lower values of $H(z\gtrsim1)$ so that the strict constraints of the CMB data on $D_M(z_*)$ are satisfied (see \cref{sec:omnipotent}).
Note that, this seesaw behavior around the Planck best fit as a consequence of the DMS20 DE density that decreases towards the past not only results in a higher $H_0$ value, but also an arguably better description of the $H(z)$ data from BAO. For the full dataset combination, the evidence for the DMS20 DE parametrization is inconclusive with respect to the $\Lambda$CDM scenario. In other words, it fits the data as well as the standard cosmological scenario while simultaneously reducing the Hubble tension.

We have carried out our discussion so far over the usual PDL crossing, i.e., the constraints on the parameter $a_m$ (p-quintessence--p-phantom crossing), whereas the DMS20 DE parametrization has two more free parameters, $\alpha$ and $\beta$. The constraints on the $\alpha$ and $\beta$ parameters can be seen in~\cref{tab:constraints} and their correlations with some other parameters on \cref{fig:all}. However, these two free parameters do not have obvious physical meanings on their own, hence we do not discuss them any further. Rather, we will focus on two physically meaningful derived parameters obtained from combinations of $a_m$, $\alpha$, and $\beta$, viz., $a_{\rm p}$ (n-quintessence--p-phantom crossing), $a_n$ (n-quintessence--p-phantom crossing), as well as $\rho_\Lambda$ (viz., minimum of $\rho_{\rm DE}$), with a focus on the sign changing energy density feature of the DMS20 DE---for the definitions of these parameters see \cref{sec:DMS20}.

To discuss the features of DMS20 DE related to negative energy densities,
we present \cref{fig:H02D,fig:DEplots,fig:DE+} showing constraints on some relevant functions and parameters for our most extensive data combination, viz., Planck+BAO+PantheonPlus\&SH0ES, and also for the Planck+PantheonPlus\&SH0ES data combination in which case the features related to negative DE density are most emphasized within DMS20 DE.

We present the two-dimensional marginalized constraints of $H_0$ versus $a_{\rm p}$ in \cref{fig:H02D} to show the correlation between the Hubble constant and the scale of the Universe at which n-quintessence--p-phantom crossing occurs, along with those of $H_0$ versus $a_n$ and $a_m$. In the Planck+PantheonPlus\&SH0ES case, we see a positive correlation between $H_0$ and $a_{\rm p}$, while no correlation is present with the other two parameters $a_n$ and $a_m$. On the other hand, once BAO data are included, the correlation between $H_0$ and $a_{\rm p}$ becomes weak,\footnote{This weakening of the correlation between $H_0$ and the time of crossing to negative densities, accompanying the lesser constraints on $a_{\rm p}$ with the addition of the BAO data, was also found within the $\Lambda_{\rm s}$CDM model~\cite{Akarsu:2021fol,Akarsu:2022typ} which has other parallelisms with DMS20 DE as noted further below in the main text; compare the leftmost panel of \cref{fig:H02D} with the bottom leftmost panel of Fig.~2 in Ref.~\cite{Akarsu:2022typ} and Fig.~8 in Ref.~\cite{Akarsu:2021fol}. Within $\Lambda_{\rm s}$CDM, the strict CMB-based constraints on $D_M(z_*)$ enforces a nonlinear degeneracy between the time of crossing and $H_0$; the shape of this degeneracy (see also Fig.~2 in Ref.~\cite{Akarsu:2021fol} and relevant discussions therein), which implies a weakened linear correlation if the crossing happens at earlier times, explains these findings for $\Lambda_{\rm s}$CDM, and it is conceivable that the situation is similar for DMS20 DE but its two more extra parameters compared to $\Lambda_{\rm s}$CDM complicates laying out an analog argumentation.} and a slight negative correlation appears between $H_0$ and $a_m$ with still no correlation between $H_0$ and $a_n$. We also note the constraints $a_n=0.675^{+0.018}_{-0.023}$ and
$a_m=0.957^{+0.016}_{-0.023}$ at $68\%$ CL for the Planck+PantheonPlus\&SH0ES case, that become $a_n=0.713^{+0.038}_{-0.076} $ and
$a_m=0.922^{+0.041}_{-0.035}$ when BAO data are included.

An $a_{\rm p}$ value less than zero would imply $\rho_{\rm DE}$ never takes negative values, whereas the constraints are $a_{\rm p}=0.357^{+0.018}_{-0.010} $ at $68\%$ CL for Planck+PantheonPlus\&SH0ES and $a_{\rm p}=0.230^{+0.066}_{-0.029} $ at $68\%$ CL when the BAO data are also included; even for the Planck+BAO+PantheonPlus\&SH0ES case, $a_{\rm p}>0$ at 99\% CL. For both data combinations, the n-quintessence--p-phantom crossing is clear in \cref{fig:DEplots} where we show the posteriors for the evolution of $\rho_{\rm DE}(z)/\rho_{c0}$ and the corresponding EoS parameters, $w_{\rm DE}(z)$. We do not show the CL contours due to the singular behavior of $w_{\rm DE}(z)$; rather, the panels are produced by drawing plots of randomly selected samples from our Markov chain Monte Carlo chains with the \texttt{fgivenx} package~\cite{Handley_2018}, i.e., the more frequent the lines, the more probable. In terms of redshift, the constraints on the time of n-quintessence--p-phantom crossing are $z_{\rm p}=1.803^{+0.069}_{-0.14}$ at $68\%$ CL for Planck+PantheonPlus\&SH0ES and $z_{\rm p}=3.7^{+0.2}_{-1.4}$ at $68\%$ CL when the BAO data are also included; the singularities of $w_{\rm DE}(z)$ are at redshifts satisfying $\rho_{\rm DE}(z)=0$, i.e., the redshifts of the n-quintessence--p-phantom crossings. Such singularities\footnote{For some examples of these types of singularities in the EoS parameter, see Refs.~\cite{Sahni:2004fb,Tsujikawa:2008uc,Zhou:2009cy,Bauer:2010wj,Sahni:2002dx,Sahni:2014ooa,Wang:2018fng,Akarsu:2019hmw,Escamilla:2021uoj,Akarsu:2019ygx,Acquaviva:2021jov,Akarsu:2022lhx,DiValentino:2017rcr,Ong:2022wrs,Akarsu:2019pvi,Akarsu:2022typ,Vazquez:2023kyx}.} mentioned in \cref{sec:DMS20,sec:omnipotent} are required from a minimally interacting (i.e., conserved) DE that changes the sign of its energy density, and are known to exist as expected within DMS20 DE~\cite{Ozulker:2022slu}.
 
It is worth noting here the striking consistence of the constraints on the redshift of crossing to negative energy densities between DMS20 DE and another model with this feature dubbed $\Lambda_{\rm s}$CDM that extends $\Lambda$CDM with one extra free parameter by replacing the usual cosmological constant with a sign-switching one~\cite{Akarsu:2021fol,Akarsu:2022typ}. In Ref.~\cite{Akarsu:2022typ}, $\Lambda_{\rm s}$CDM was analysed with a data combination referred as CMB+Pan+$M_B$ and yielded a constraint on the crossing redshift (denoted with $z_\dagger$ in there) $z_\dagger=1.78^{+0.14}_{-0.18}$ along with $H_0=72.38^{+0.98}_{-1.10}$ km/s/Mpc and $S_8=0.785\pm 0.012$ at $68\%$ CL; these results are to be compared with the constraints $z_{\rm p}=1.803^{+0.069}_{-0.140}$, $H_0=73.49\pm 0.98$ km/s/Mpc and $S_8=0.803\pm 0.011$ for our Planck+PantheonPlus\&SH0ES case due to the parallelisms of the datasets used in both papers explained below. The data combination CMB+Pan+$M_B$ in Ref.~\cite{Akarsu:2022typ} consists of the Pantheon sample of SNe Ia~\cite{Pan-STARRS1:2017jku} along with Cepheid calibrated SNe Ia absolute magnitude measurement of SH0ES~\cite{Camarena:2021jlr} and the same Planck CMB data used in this work. The part of this data combination related to SNe Ia can be considered as an earlier version of the PantheonPlus\&SH0ES dataset used in the present work, and since the CMB data is the same for both papers, the CMB+Pan+$M_B$ data combination of Ref.~\cite{Akarsu:2022typ} contains very similar information with the Planck+PantheonPlus\&SH0ES here. Moreover, when the same BAO data used here was added to their data combination CMB+Pan+$M_B$ (referred as CMB+Pan+BAO+$M_B$), the constraints on the redshift of crossing shifts to higher values, i.e., $z_\dagger=2.36\pm0.28$ accompanied by worsened values of $H_0=69.48^{+0.48}_{-0.55}$ km/s/Mpc and $S_8=0.813\pm 0.010$ parallel to the Planck+BAO+PantheonPlus\&SH0ES case in here which yields $z_{\rm p}=3.7^{+0.2}_{-1.4}$, $H_0=70.05\pm 0.64$ km/s/Mpc and $S_8=0.823\pm 0.011$.

Finally, we decompose the DE density into a cosmological constant and non-negative DE$^+$ density as described in~\cref{subsec:decomposition}, viz. we write $\rho_{\rm DE}(a)=\rho_\Lambda+\rho_{\rm DE^+}(a)$, where the minimum of the DMS20 DE density, $\rho_\Lambda=\rho_{\rm DE}(a_{\rm min})$, can attain negative values and $\rho_{\rm DE^+}\geq0$. The top panels of~\cref{fig:DEplots} show that we can safely assume $a_{\rm min}=0$ in line with the results of Ref.~\cite{DiValentino:2020naf}; this yields $\rho_\Lambda/\rho_{\rm c0}=-5.1^{+1.0}_{-1.0}$ (along with $\rho_{\rm DE^+}/\rho_{\rm c0}=5.8^{+1.0}_{-1.0}$) at 68\% CL for Planck+PantheonPlus\&SH0ES and $\rho_\Lambda/\rho_{\rm c0}=-1.5^{+0.8}_{-0.6}$ (along with $\rho_{\rm DE^+}/\rho_{\rm c0}=2.22^{+0.60}_{-0.83}$) at 68\% CL when the BAO data is also included. In both cases, $\rho_\Lambda<0$ at 95\% CL indicating AdS vacua is preferred within DMS20 DE by our CMB and SNe Ia based datasets albeit the weakening of this preference with the inclusion of the BAO data. In \cref{fig:DE+}, we present the posteriors for $\rho_\Lambda$ along with the corresponding DE$^+$ densities and EoS parameters; we use line plots instead of showing CL contours for visual homogeneity with \cref{fig:DEplots}. We notice the DE$^+$, on top of the negative cosmological constant (viz. $\rho_\Lambda<0$), starts as a p-phantom DE with an EoS parameter equal to $-4/3$ in the early universe by construction and then evolves as the Universe expands transforming into a DE that exhibits a positive cosmological constant-like behavior [viz. $\rho_{\rm DE^+}(a)$ remains almost constant] in the late universe, namely, for $z\lesssim1$. It is very interesting that $\rho_{\rm DE^+}(a)$ achieves its almost constant behavior for $z\lesssim1$ by a slightly oscillatory behavior of its corresponding EoS parameter around PDL in the late universe ($z\lesssim1$); i.e., it is above PDL (p-quintessence) for $a_n<a<a_m$ and remains p-phantom everywhere else. We also note that, the future phantom behavior of DMS20, i.e., $w_{\rm DE^+}(a\to\infty)=w_{\rm DE}(a\to\infty)=-2$, implies a future singularity for the Universe by means of a ``big rip"~\cite{McInnes:2001zw,Caldwell:2003vq} for the choice of our priors. However, for some regions in the full parameter space of DMS20, a singularity by means of a ``big crunch" is also possible after an accelerated era due to a DE density that attains negative values in the future, e.g., see the dark orange plot in \cref{fig:demo}.

\section{Conclusion}
\label{sec:conc}

In this paper, we introduce omnipotent DE models: the family of DE models whose nonmonotonically evolving energy densities can cross between negative and positive values and correspond to EoS parameters that feature PDL crossings (at the times their density manifests nonmonotonicity) and even evolve to attain all values by means of singularities (at the times their densities changes sign). These features characterizing omnipotent DE, if they were to be confirmed, would have far reaching consequences for fundamental physics; however, they may be more benign then one may initially think. The usual energy conditions they obviously violate, do not have \textit{a priori} justifications~\cite{Curiel:2014zba,Epstein:1965zza,Wald:1991xn,Visser:1999de}. Although an individual component like the omnipotent DE apparently violates some of the energy conditions, the total energy density of the Universe always satisfies the energy conditions except the strong energy condition, which is needed for the accelerating Universe. Also the phenomenological omnipotent DE characteristics can possibly be modelled in modified theories of gravity, for instance, in a massive Brans-Dicke (BD) theory~\cite{Uehara:1981nq,Faraoni:2009km,Boisseau:2010pd,Akarsu:2019pvi} all behaviors in \cref{tab:omni} are present
except the negative-CC which can also be modelled with the BD scalar field potential having an AdS minimum. Moreover, from the point of view of observational tensions that cannot be satisfactorily addressed by the canonical/simple extensions of $\Lambda$CDM, these energy condition violating features are strongly motivated; they have been repeatedly found in studies reconstructing cosmological functions~\cite{Bonilla:2020wbn,Sahni:2014ooa,Aubourg:2014yra,Wang:2018fng,Poulin:2018zxs,Escamilla:2021uoj,Escamilla:2023shf}, and models whose parameter space allow them, have shown promising success in alleviating the observational discordance within $\Lambda$CDM by occupying parts of their parameter space that exhibit these features~\cite{Visinelli:2019qqu,Sen:2021wld,Calderon:2020hoc,Sahni:2014ooa,DiValentino:2020naf,Akarsu:2019hmw,Dutta:2018vmq,Akarsu:2021fol,Akarsu:2022typ,Akarsu:2019ygx,Acquaviva:2021jov,Akarsu:2022lhx,Vazquez:2023kyx}---one such model is DMS20.

DMS20 parametrization of the DE density introduced in Ref.~\cite{DiValentino:2020naf} can be obtained by Taylor expanding the DE density to third order around a presupposed PDL crossing; it is an embodiment of omnipotent DE. While DMS20 was proposed as a means of investigating the existence of a PDL crossing in the DE EoS parameter, and its observational success when confronted with various combinations of CMB, SNe Ia and BAO data was attributed to a PDL crossing being favored by these observations, it was also found that the constrained parameter space of DMS20 yields a DE density that attains negative values in the past. In this paper, we reanalyze DMS20, in the context of omnipotent DE, with an enhanced focus on its negative density feature. We first give a preliminary analysis of the model to present the background phenomena corresponding to different regions of its parameter space; and, we put forth an interpretation of the DMS20 DE as a combination of a negative valued $\Lambda$ and a non-negative dynamical DE. Then, we confront the model with observational data updating and extending those of Ref.~\cite{DiValentino:2020naf}.
With regard to the tensions of $\Lambda$CDM, the DMS20 model relaxes the $H_0$ tension for all of our data combinations with a promising negative correlation between $H_0$ and $S_8$ parameters; and, for our Planck+PantheonPlus\&SH0ES data combination, its excellent agreement with the SH0ES $H_0$ measurement is accompanied by an excellent agreement with the $t_0$ estimation from GCs. In comparison, the widely popular EDE models that are proposed to address the cosmological tensions, while able to alleviate the $H_0$ tension, exacerbate the $S_8$ discrepancy and tend to predict lower $t_0$ values compared to the estimation from GCs~\cite{Poulin:2023lkg}.
Our results indicate that DMS20 DE alleviates observational discrepancies of $\Lambda$CDM by means of a coalition between its negative density and PDL crossing features; in particular, the scale factor that energy density changes sign, $a_{\rm p}$, is positively correlated with $H_0$, a finding directly parallel with the results of the $\Lambda_{\rm s}$CDM model's analyses in Refs.~\cite{Akarsu:2021fol,Akarsu:2022typ}. Moreover, when DMS20 is decomposed to a non-negative dynamical DE on top of a cosmological constant, our data combinations that include the SH0ES calibration of SNe Ia suggest a negative cosmological constant at 95\% CL. It is worth noting the agreement of this result with the recent claim that PantheonPlus dataset itself demands negative DE density at high redshifts~\cite{Malekjani:2023dky}. Of course, the results discussed in this paragraph are specific to the DMS20 parametrization rather than general omnipotent DE models, and hint that omnipotent features (negative and nonmonotonic energy density) might be necessary to address the cosmological tensions satisfactorily.

We motivate the omnipotent DE family based on the observational success of DE models that incorporate some or all of its phenomenological features and the fact that observational reconstructions have consistently found these behaviors~\cite{Visinelli:2019qqu,Sen:2021wld,Calderon:2020hoc,Sahni:2014ooa,DiValentino:2020naf,Akarsu:2019hmw,Dutta:2018vmq,Akarsu:2021fol,Akarsu:2022typ,Akarsu:2019ygx,Acquaviva:2021jov,Akarsu:2022lhx,Vazquez:2023kyx,Bonilla:2020wbn,Sahni:2014ooa,Aubourg:2014yra,Wang:2018fng,Poulin:2018zxs,Escamilla:2021uoj,Escamilla:2023shf}---although, these findings may just be artifacts of methodology, and it is not yet clear if some of the data that indicate deviations from $\Lambda$CDM are results of systematic errors or statistical flukes. Also, a negative DE in the past, a major component of omnipotent DE, has theoretical motivation due to the better compatibility of a negative $\Lambda$ (compared to a positive one) with string theory and string-theory-motivated supergravities~\cite{Maldacena:1997re,Bousso:2000xa}. 
The high number of features of omnipotent DE, while allows a good description of the available cosmological data, could be challenging to achieve without introducing a high number of extra free parameters on top of $\Lambda$CDM, e.g., the DMS20 parametrization introduces three. However, it is encouraging that, for instance, massive Brans-Dicke theory extensions of the $\Lambda$CDM model~\cite{Uehara:1981nq,Faraoni:2009km,Boisseau:2010pd}, which has only one extra free parameter (viz., the Brans-Dicke parameter $\omega$), incorporates all behaviors of omnipotent DE except negative-CC. While the motivation behind the omnipotent DE family is mostly phenomenological, its current success in alleviating the tensions, especially if it is to be clinched by future probes, motivates looking for underlying mechanisms of its quirky features.

\begin{acknowledgments}
\"{O}.A. acknowledges the support by the Turkish Academy of Sciences in scheme of the Outstanding Young Scientist Award  (T\"{U}BA-GEB\.{I}P). E.~D.~V. is supported by a Royal Society Dorothy Hodgkin Research Fellowship.  R.~C.~N. thanks the CNPq for partial financial support under the Project No. 304306/2022-3. E.~\"{O}.~acknowledges the support by The Scientific and Technological Research Council of Turkey (T\"{U}B\.{I}TAK) in scheme of 2211/A National PhD Scholarship Program. A.~A.~S. acknowledges the funding from SERB, Govt of India under the research Grant No. CRG/2020/004347. A.~A.~S. and S.~A.~A. acknowledges the use of High Performance Computing facility Pegasus at IUCAA, Pune, India. This article is based upon work from COST Action CA21136 Addressing observational tensions in cosmology with systematics and fundamental physics (CosmoVerse) supported by COST (European Cooperation in Science and Technology).
\end{acknowledgments}

\bibliography{bibliorev}
\end{document}